\documentclass{aa}
\pdfoutput=1 
\usepackage{graphicx}
\usepackage{natbib}
\usepackage{pifont}
\usepackage{lscape}
\usepackage{epsfig}
\usepackage{amsmath}
\usepackage{longtable}
\usepackage{ltcaption}
\usepackage{dblfloatfix}
\usepackage{txfonts}
\usepackage[colorlinks=true,citecolor=blue]{hyperref}
\begin{document} 

   \title{An innovative blazar classification based on radio jet kinematics}

  \titlerunning{Kinematic blazar classification}
   \author{O. Hervet
          \inst{1,2}
          ,
          C. Boisson
          \inst{1},
          \and
          H. Sol
          \inst{1}
          }

   \institute{LUTH, Observatoire de Paris, PSL, CNRS, Université Paris Diderot, 5 Place Jules Janssen, 92190 Meudon, France\\
   \and Santa Cruz Institute for Particle Physics and Department of Physics, University of California at Santa Cruz, Santa Cruz,
CA 95064, USA\\
              \email{ohervet@ucsc.edu}
             }

   \date{Received January 13, 2016; accepted April 30, 2016}

 
  \abstract
  {Blazars are usually classified following their synchrotron peak frequency ($\nu F(\nu)$ scale) as high, intermediate, low frequency peaked BL Lacs (HBLs, IBLs, LBLs), and flat spectrum radio quasars (FSRQs), or, according to their radio morphology at large scale, FR~I or FR~II. However, the diversity of blazars is such that these classes seem insufficient to chart the specific properties of each source.}
  {We propose to classify a wide sample of blazars following the kinematic features of their radio jets seen in
very long baseline interferometry (VLBI).}
  {For this purpose we use public data from the MOJAVE collaboration in which we select a sample of blazars with known redshift and sufficient monitoring to constrain apparent velocities. We selected 161 blazars from a sample of 200 sources. We identify three distinct classes of VLBI jets depending on radio knot kinematics: class I with quasi-stationary knots, class II with knots in relativistic motion from the radio core, and class I/II, intermediate, showing quasi-stationary knots at the jet base and relativistic motions downstream.}
  {A notable result is the good overlap of this kinematic classification with the usual spectral classification; class I corresponds to HBLs, class II to FSRQs, and class I/II to IBLs/LBLs. We deepen this study by characterizing the physical parameters of jets from VLBI radio data. Hence we focus on the singular case of the class I/II by the study of the blazar BL Lac itself. Finally we show how the interpretation that radio knots are recollimation shocks is fully appropriate to describe the characteristics of these three classes.}
  {}


   \maketitle
%

\section{Introduction}

Radio very long baseline interferometry (VLBI) can achieve angular resolutions that are smaller than milliarcseconds. This has revealed the structure of active galactic nuclei (AGN) jets in close contact with the radio core, reaching subparsec scale on the sky plan for the nearest sources. 
Radio VLBI jets present bright knots with origin and properties still poorly understood.
Some of these knots are observed to have relativistic motions in the jet, which are identifiable by their superluminal apparent velocities. Knot properties, such as size, apparent velocity, and luminosity, are used in various studies to constrain the Doppler factor of the non-thermal emission zone \citep[e.g.][]{Lahteenmaki_1999,Jorstad_2005,Hovatta_2009}.

Recent works show that flat spectrum radio quasars (FSRQs) have on average radio apparent velocities higher than BL Lacs \citep{Lister_2013}. Some BL Lacs presenting quasi-stationary knots, however, are known to exhibit variabilities on very short timescale and show significant flux level at very high energies, de facto requiring very high Lorentz factors.
Current scenarios attempting to explain this phenomenon propose that jets could mark a strong deceleration near the core, implying that the variable emission zone is situated upstream from the radio knots \citep{Georganopoulos_2003, Piner_2004}. Such deceleration near the core is hardly compatible with the size of radio jets observed reaching up to Mpc, thus requiring a strong kinetic power preserved at large scale. Moreover some sources show relativistic motions far from the core and quasi-stationary knots upstream, such as the M 87 radio galaxy \citep{Kovalev_2007,Giroletti_2012}, which is not consistent with the interpretation of a strong flow deceleration.
Another scenario, developed by \cite{Marscher_1985} and deepened by \cite{Meier_2013}, considers a strong stationary recollimation shock at the jet base that is responsible for the high energy emission of the source.
Such a shock could be associated with either  the radio core or with a stationary knot at the jet base \citep{Cohen_2014}, implying that the knot motion is decorrelated from the underlying flow velocity.

In this paper we approach various questions regarding the singular kinematics of radio knots and its relevance in the current blazar classification scheme with the study of a sample of 161 sources.
In Section \ref{Section::Relevance of a kinematical blazar classification}, we discuss criteria that effectively link the kinematic properties of the radio knots to the spectral class of blazars. In Section \ref{Section::Jets physical parameters}, we develop a method allowing the deduction of some jet physical parameters from the radio knots properties. Hence in Section \ref{Section::Study of an intermediate blazar, the BL Lac case}, we study valuable information about the nature of the knots that is given by the intermediate blazar BL Lac itself.
 Finally, in Section \ref{Section::Discussion and interpretation}, a qualitative discussion highlighting the interpretation of knotty structures by multiple recollimation shocks is presented.

In the following, we use a cosmology with $H_0 = 71$ km.s$^{-1}$.Mpc$^{-1}$, $\Omega_M = 0.27,$ and $\Omega_V = 0.73$. 

\section{Relevance of a kinematical blazar classification}
\label{Section::Relevance of a kinematical blazar classification}

The average maximal apparent velocity of the radio knots of FSRQs  is slightly higher than that of BL Lacs, nevertheless there is a strong overlap that prevents an efficient discrimination between these different classes of objects based on this criterion alone \citep{Lister_2013}. These authors also mentioned that BL Lacs are more likely to show quasi-stationary knots than FSRQs.

In order to investigate the role of apparent velocities and quasi-stationary features in blazar jets more deeply,  we focus on the largest well-monitored sample of AGN radio VLBI jets: the MOJAVE\footnote{Monitoring Of Jets in Active galactic nuclei with VLBA Experiments, \url{http://www.physics.purdue.edu/MOJAVE/}} collaboration source catalogue.
From the original sample of 200 AGN, we selected
\begin{itemize}
\item blazars with known redshift to access the projected real size of jets and components.
\item blazars that are sufficiently monitored to allow an estimation of apparent velocities by MOJAVE.
\end{itemize}

This leads to a sample of 161 sources, which can be classified by distinct kinematic structures of jets.
Indeed all HBLs in the sample have radio knot apparent velocities $\beta_{app} < 2$ (in c units), when knots in other blazar types exhibit strong relativistic motions. Distinguished by minimal and maximal apparent velocities, three classes are identified:

\begin{itemize}
\item{Class I}: Blazars with quasi-stationary knots or with "low" apparent velocities (max($\beta_{app}$)$ < 2$): 25 sources.

\item{Class II}: Blazars with knots in relativistic motion from the jet base (max($\beta_{app}$) $\geq$ 2): 99 sources.

\item{Class I/II}: Blazars with quasi-stationary knots close to the jet base (min($\beta_{app}$) $\leq$ 1) and in relativistic motion downstream (max($\beta_{app}$) $\geq$ 2): 37 sources.
\end{itemize}

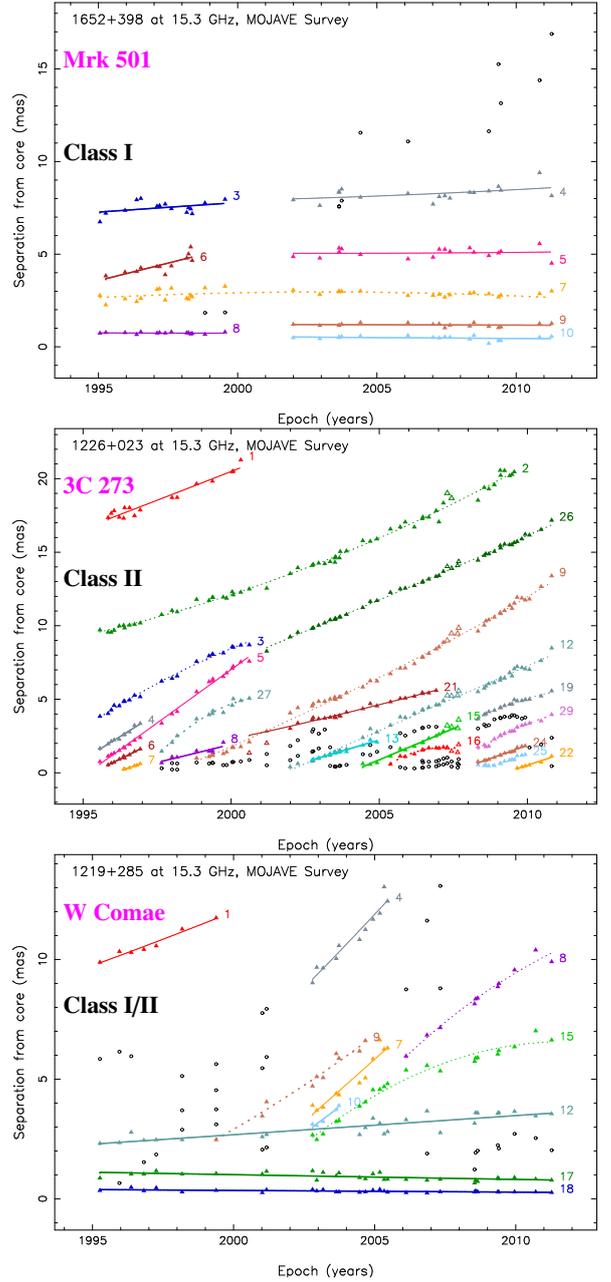
\begin{figure}[h]
      \centering \includegraphics[width=7.7cm]{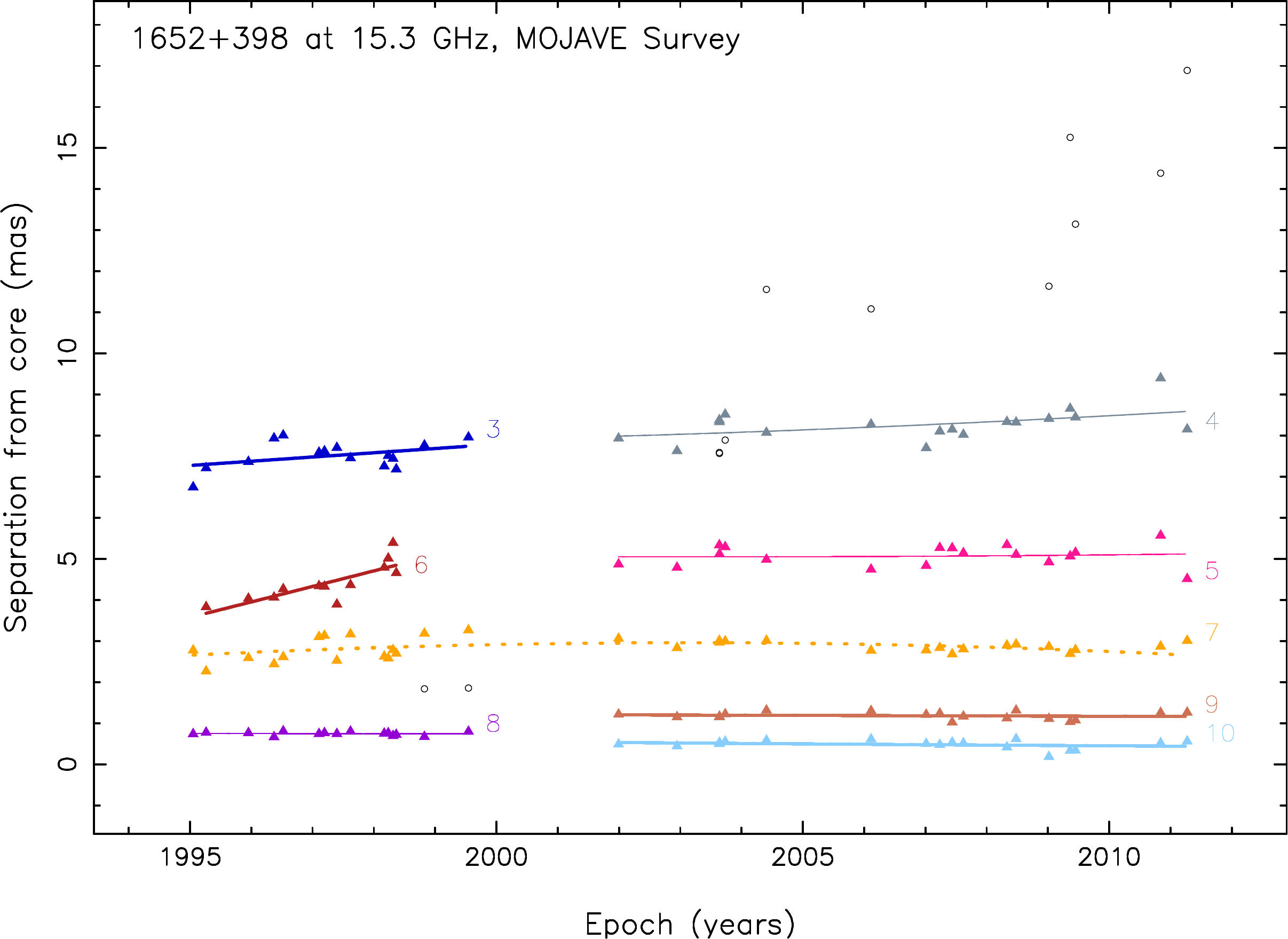}
      \put(-200,135){\color[rgb]{1,0,1}\makebox(1,0)[lb]{\smash{\textbf{Mrk 501}}}}
      \put(-200,100){\color[rgb]{0,0,0}\makebox(1,0)[lb]{\smash{\textbf{Class I}}}}

      \centering \includegraphics[width=7.7cm]{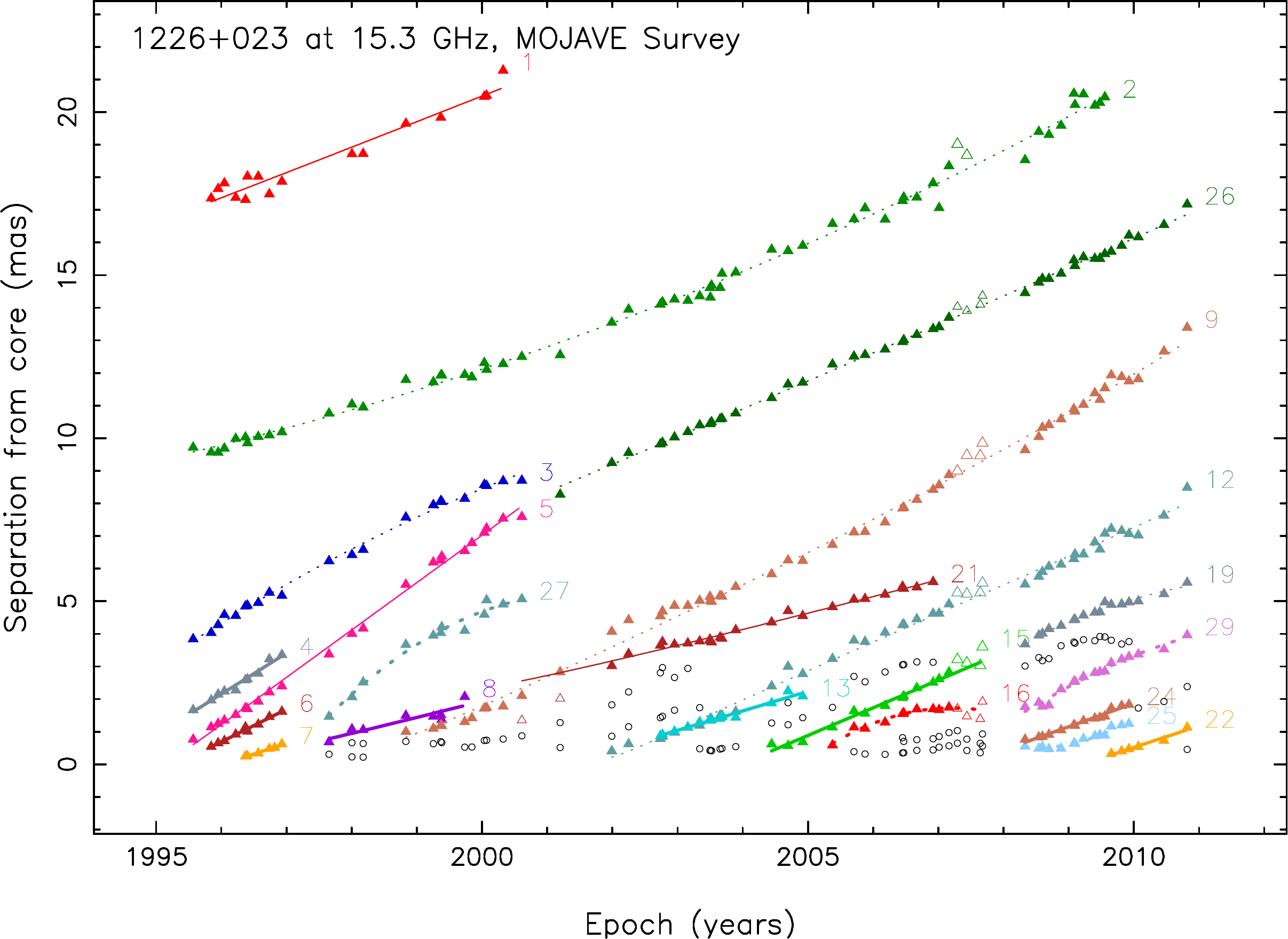}
      \put(-200,135){\color[rgb]{1,0,1}\makebox(1,0)[lb]{\smash{\textbf{3C 273}}}}
      \put(-200,100){\color[rgb]{0,0,0}\makebox(1,0)[lb]{\smash{\textbf{Class II}}}}

   \centering \includegraphics[width=7.7cm]{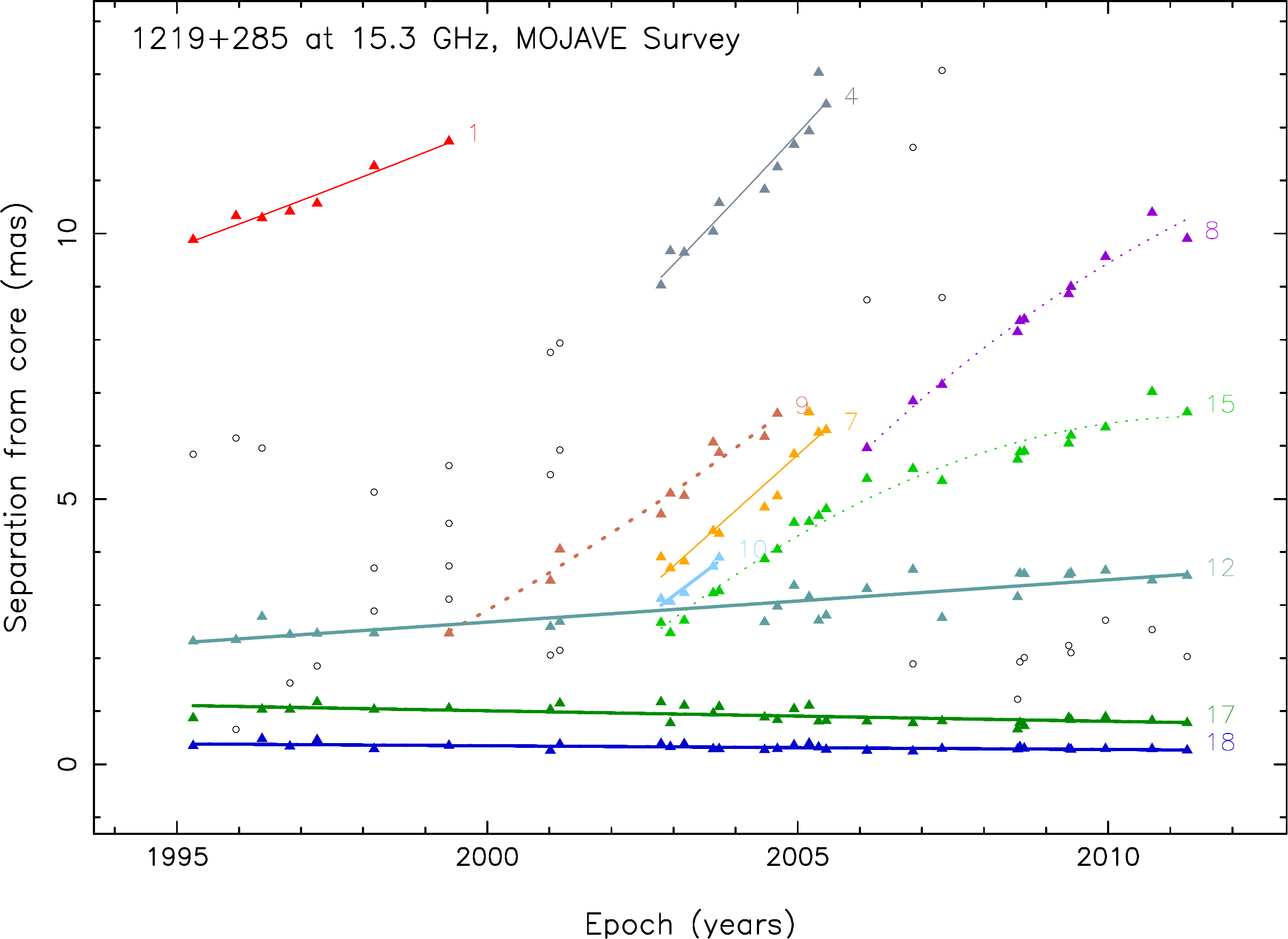}
   \put(-200,135){\color[rgb]{1,0,1}\makebox(1,0)[lb]{\smash{\textbf{W Comae}}}}
   \put(-200,100){\color[rgb]{0,0,0}\makebox(1,0)[lb]{\smash{\textbf{Class I/II}}}}
   \caption{Temporal evolution of knot-core distances for three blazars representative of the three kinematic classes I, II and I/II, selected from the MOJAVE database.}
   \label{Fig::classes_cinematiques}
\end{figure}

These criteria highlight the various kinematic structures shown in Figure \ref{Fig::classes_cinematiques}. The name of the last class,  I/II, is chosen considering its hybrid nature between classes I and II.

The presence of non-trivial kinematic behaviours in some sources leads us to clarify some aspects of the classes assignation.
Sources with relativistic apparent velocities from their jet base are classified as II, even if they show one or several quasi-stationary knots downstream. The main criterion defining the class II is the ability of producing relativistic knots very close to the core. The sources B2 0202+31, S4 0917+44, 4C +29.45, PKS 1655+077, PHL 5225, CTA 102, 3C 418, PKS 2134+004, 4C +31.63, and also 3C 454.3, as noticed by \cite{Jorstad_2013}, show this feature.
Also, sources with knots close to the core without associated apparent velocities but with relativistic velocities downstream are classified as I/II. The sources Ap Librae, TXS 0730+504, and TXS 2005+403 show this feature.

Some biases can also affect this classification. 
A lack of monitoring of a source I/II increases the risk of detecting only velocities from quasi-stationary knots, leading to confusion with class I. This effect does not significantly impact  the class II sample, as described above; only $10\%$ of this sample presents quasi-stationary components.
Finally, all sources close to the discriminating apparent velocity max($\beta_{app}$) = 2 have a higher risk to be misclassified. To estimate this bias, we count the borderline sources in the interval max($\beta_{app}$) $\in [1.5,2.5]$. They represent $28\%$ of class I, $3\%$ of class I/II and $4\%$ of class II. This effect is thus potentially significant for class I, which is not surprising because an isotropic repartition of velocities between 0 and 2 c gives $25\%$ of sources in this interval.

\subsection{Overlap with spectral classes}

A first test of the consistency of this classification is to check its overlap with the standard spectral classes. Blazars are divided in three spectral classes: FSRQs, LBLs/IBLs, and HBLs, which correspond to 125, 23, and 5 sources in our sample, respectively (Table \ref{table::recoupement_classe_spectrale}). Only sources with well-defined spectral classes are taken into account. The HBL class, which is  less luminous in radio, is unfortunately under-represented in the MOJAVE database.

\begin{table}[h]
\caption{Overlap of the kinematic classification with spectral classification.}
\label{table::recoupement_classe_spectrale}
\centering
\begin{tabular}{lcccc}
\hline\hline
\noalign{\smallskip}
Spectral class & number & Class I & Class I/II & Class II \\
\hline
\noalign{\smallskip}
HBLs & 5 & $100\%$ & $0\%$ & $0\%$ \\
IBLs/LBLs & 23 & $32\%$ & $56\%$ & $12\%$\\
FSRQs & 125 & $8\%$ & $16.5\%$ & $75.5\%$ \\
\hline
\end{tabular}
\end{table}

All five blazars referenced as HBLs belong to class I. The IBLs and LBLs belong predominantly to class I/II, but show a significant spread in the other classes. The strong presence of IBLs/LBLs in class I ($32\%$) could be due to the observational bias previously expressed that the less a source is monitored in VLBI, the more chances there are to interpret the source as a class I. The FSRQs are clearly grouped in class II ($75\%$) with a more pronounced spread in class I/II than I, as expected.

This new kinematic classification of blazars proposed here shows a strong link with the standard spectral classification. The discovery of such a link for IBLs and LBLs may provide a valuable key to the understanding of transitions between different classes.

\subsection{Overlap with large-scale radio jets}

\begin{figure}[h]
   \centering \includegraphics[width=9.0cm]{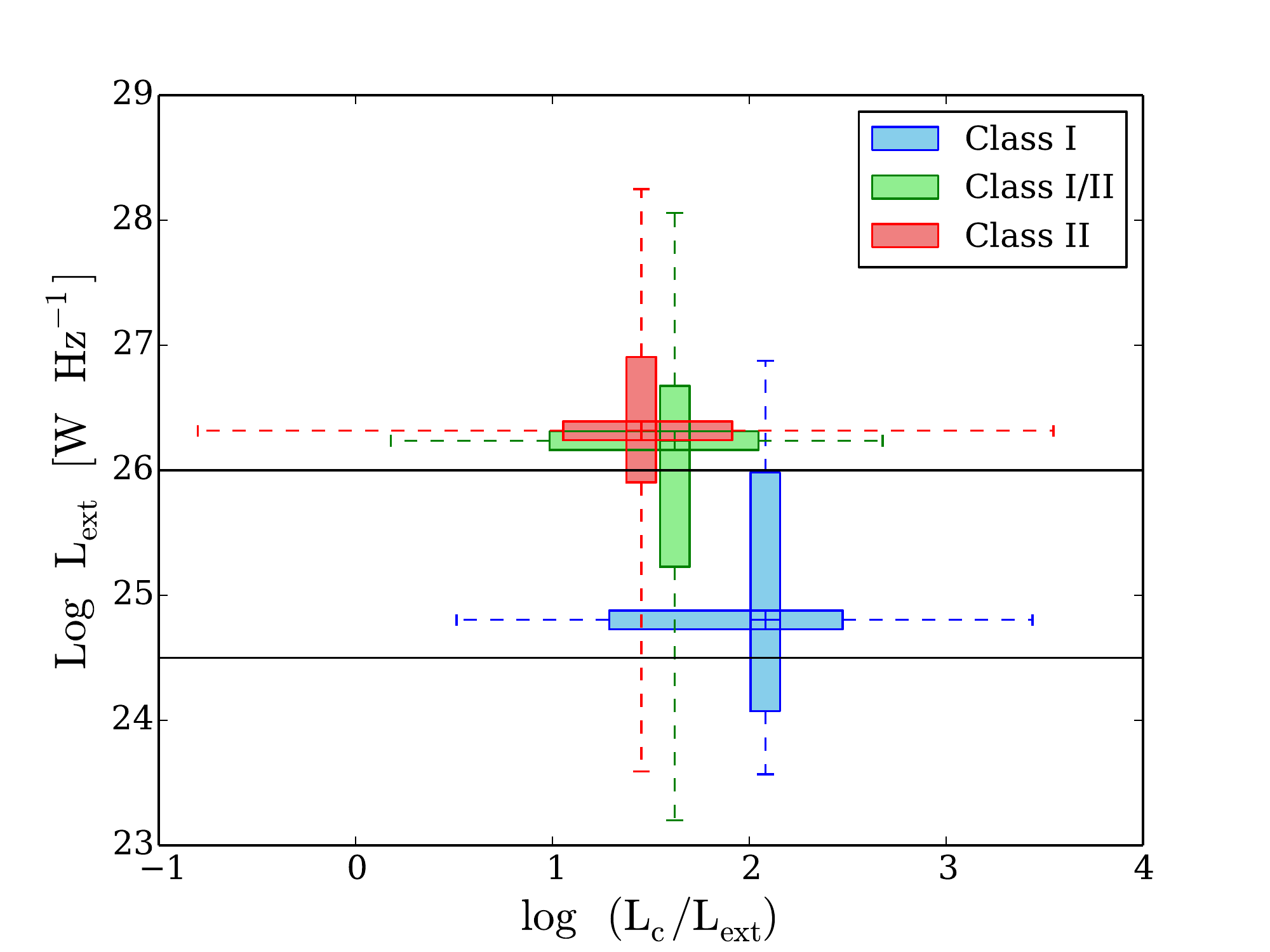}
   \put(-220,40){\color[rgb]{0,0,0}\makebox(1,0)[lb]{\smash{\textbf{FR I}}}}
   \put(-220,75){\color[rgb]{0,0,0}\makebox(1,0)[lb]{\smash{\textbf{FR I/II}}}}
   \put(-220,150){\color[rgb]{0,0,0}\makebox(1,0)[lb]{\smash{\textbf{FR II}}}}
   \caption{Box plots of extended jets radio luminosities vs. core-extended jet luminosities ratio for the three kinematic classes. Boxes are delimited by the first and last quartiles, middle lines are medians, and dashed lines account for the maximum size
of distributions. Horizontal black lines delimit regions where the luminosities of extended jets are typical of FR~I or FR~II morphologies.}
   \label{fig::R_vs_l_ext}
\end{figure}

Large-scale, radio-loud AGN morphologies, FR~I and FR~II, are strongly dependent on extended jets luminosities in radio. Such a link has been highlighted from radio jet samples by \cite{Landt_2006} and extended by \cite{Kharb_2010}. 
A large number, 118 blazars, of our sample belong to the Kharb sample. This allows us to check the eventual links between VLBI kinematic classes and the large-scale morphologies.
The distributions of 1.4 GHz extended radio luminosities of sources and their core/extended jet ratio are plotted in Figure \ref{fig::R_vs_l_ext} for the three kinematic classes. 
As in \cite{Kharb_2010}, the solid lines indicate FR~I-FR~II divide with an undifferentiated area FRI/II between (FR~II have extended luminosities $L_{ext} \geq 10^{26}$ W.Hz$^{-1}$ and FR~I $L_{ext} \leq 10^{24.5}$ W.Hz$^{-1}$).



Classes I/II and II are mostly in the  FR~II domain with strong extended radio luminosities, contrary to class I, which present a median extended luminosity that is almost two orders of magnitude lower than the other classes.
This could suggest that large-scale radio jet properties are more affected by the VLBI kinematic difference between class I and the other classes than between classes I/II and II.
Their median values, favouring a continuity of extended jet luminosities and extended-core luminosity ratios, are expressed in Table \ref{table::med_FR}.

\begin{table}[h]
\caption{Medians of the extended radio jets luminosities ($\log$ [W.Hz$^{-1}$]) and of the luminosities ratio core/extended for the three VLBI kinematic classes.}
\label{table::med_FR}
\centering
\begin{tabular}{lccc}
\hline\hline
\noalign{\smallskip}
 & nb & $\log (L_{ext})$ & $\log (L_c/L_{ext})$ \\
\hline
\noalign{\smallskip}
Class I & 11 & 24.80 & 2.08 \\
Class I/II & 28 & 26.24 & 1.62 \\
Class II & 78 & 26.32 & 1.45 \\
\hline
\end{tabular}
\end{table}

Knowing that FR~II AGN have much more powerful jets than FR~I AGN, this trend indicates that the extended jet power follows the VLBI kinematic classification; however, the wide dispersion made this link weak for a source-by-source study.
This highlights a potentially different process in jet propagation from the class I sources, which differentiates them from the other populations.

\section{Jet physical parameters}
\label{Section::Jets physical parameters}

Radio VLBI measures, such as apparent velocities, sizes, and luminosities of knots \citep{Lister_2013}, allow us to deduce jet physical parameters. Thus, the three kinematical behaviours defined above can be now characterised following physical criteria.

\subsection{Doppler factors, Lorentz factors, and angles with the line of sight}
\label{Section::delta_Gamma_alpha}

Apparent velocities of radio knots are usually associated with the Doppler factor of underlying beam flow \citep{Daly_1996,Jorstad_2005,Onuchukwu_2013,Hervet_2015}. However, knots of class I sources present quasi-stationary motions that are incompatible with the high Doppler and Lorentz factors suggested by the fast variability of HBLs. We thus only associate apparent velocities and Doppler factors with the sources belonging to classes I/II and II.  For each source we assume that the maximum detected apparent velocity is representative of the flow velocity of jets.

The Doppler factor $\delta$ depends on the apparent velocity $\beta_{app}$, but also on the angle $\theta$ between the direction of the jet and  line of sight,
\begin{equation}
\delta (\theta,\beta_{app}) = \sqrt{1- \left(\frac{\sin \theta}{\beta_{app}} + \cos \theta \right)^{-2}} \left(1+\frac{\beta_{app}}{\tan \theta}\right).
\label{eq::dopp_bapp}
\end{equation}
We simulate the radio detection probability on this angle and the maximal measured apparent velocity $P(\theta,\beta_{app})  $ to define realistic $\theta$ angles for the selected sources. For each source, the angle $\theta$ is chosen where the probability function is at maximum.

This probability function can be seen as the combination of
\begin{itemize}
\item[$\bullet$] the intrinsic angle distribution of jets projected on the sky plane $P_{proj}(\theta)$;
\item[$\bullet$] the source detection probability depending on the relativistic beaming $P_{beam}(\theta,\beta_{app})$.
\end{itemize}


Considering an isotropic distribution of AGN jet directions, their angles to the line-of-sight  $\theta \in [\theta_1, \theta_2]$, $(\theta_1 < \theta_2)$ is proportional to the annulus area of the solid angle described by $\theta_2 - \theta_1$.
So we can express the distribution $P_{proj}(\theta_n)$ as
\begin{equation}
P_{proj}(\theta_n) \propto \left[ \cos \theta_{n-1} - \cos \theta_{n} \right].
\end{equation}

This distribution favours large angles to the line of sight with a maximum at $\theta = 90^\circ$. Hence, if the AGN jets where not relativistic, one should observe many more misaligned radio galaxies than blazars. 

However, jets are highly relativistic and blazars are strongly beamed, so we have to quantify their overexposure.
Knowing that the jet radio flux $F$ comes from synchrotron radiation amplified by Doppler boosting, and that the associated radio spectrum is relatively flat, we have the relation $F \propto \delta^3$.
Hence, using the equation \ref{eq::dopp_bapp}, we can express the source detection probability depending on the relativistic beaming,
\begin{equation}
P_{beam}(\theta,\beta_{app}) \propto \sqrt{\delta^3(\theta,\beta_{app})} .
\end{equation}

With the product of $P_{proj}(\theta)$ and $P_{beam}(\theta,\beta_{app})$, the general probability can be expressed as a function of a jet direction,
\begin{equation}
P(\theta,\beta_{app}) \propto P_{proj}(\theta) \times P_{beam}(\theta,\beta_{app}).
\end{equation}

\begin{figure}[h]
\centering \includegraphics[width=9cm]{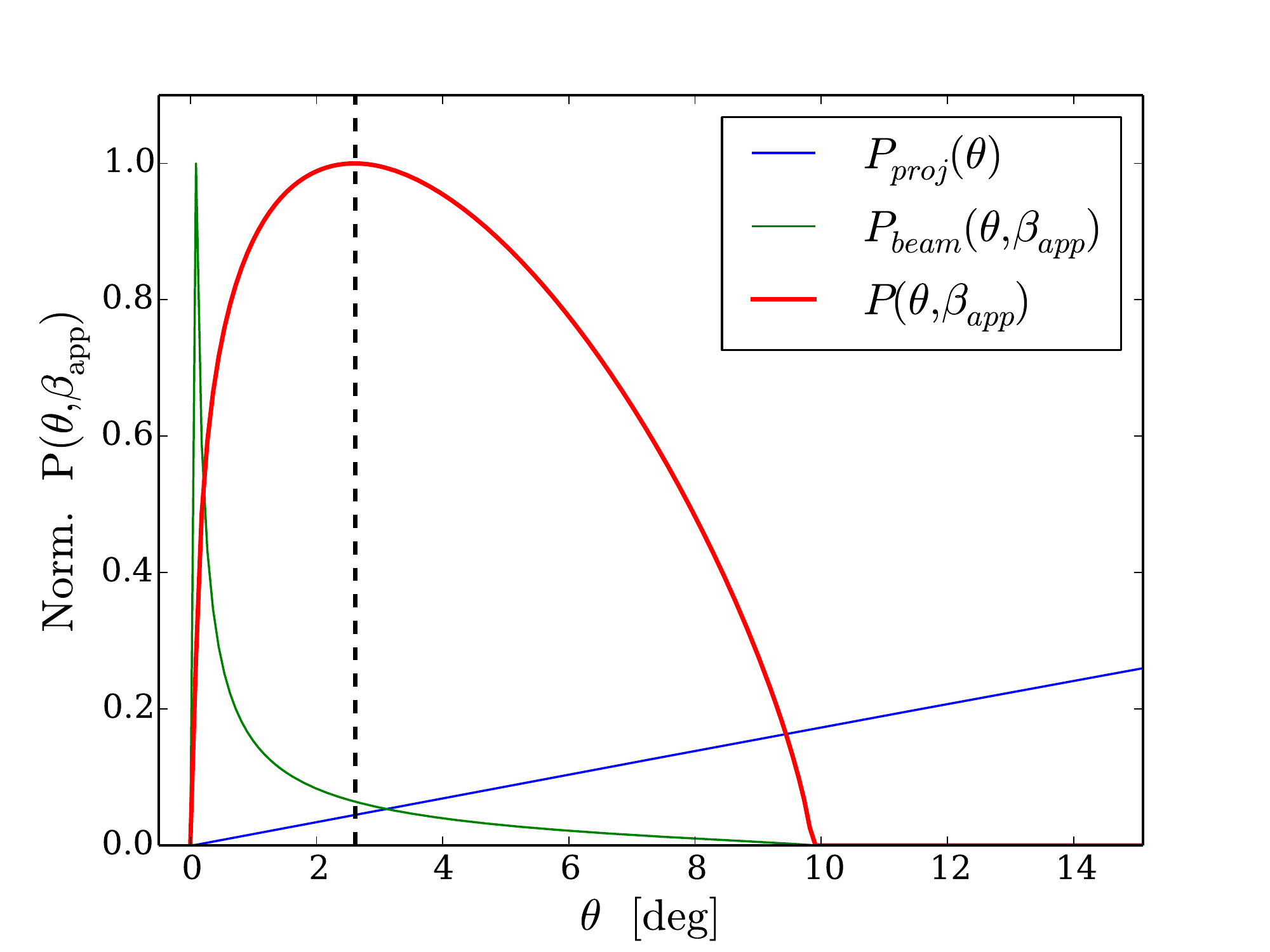}
\caption{Normalised probability of the jet angle with the line of sight for a given apparent velocity. The probability function shown here is based on the calculated median value of class II sources maximal apparent velocities, namely $\beta_{app} = 11.6$. The angle with a maximal probability, indicated with a dotted line, corresponds to $\theta = 2.6$ deg.}
\label{fig::repartition_totale}
\end{figure}

This probability depends
only on the angle to the line of sight, where $\beta_{app}$ is an observational constraint. Thus for each source of kinematic classes I/II and II, we can determine the most probable angle of the jet with the line of sight, as shown in Figure \ref{fig::repartition_totale}.

\begin{figure}[h]
\centering \includegraphics[width=9cm]{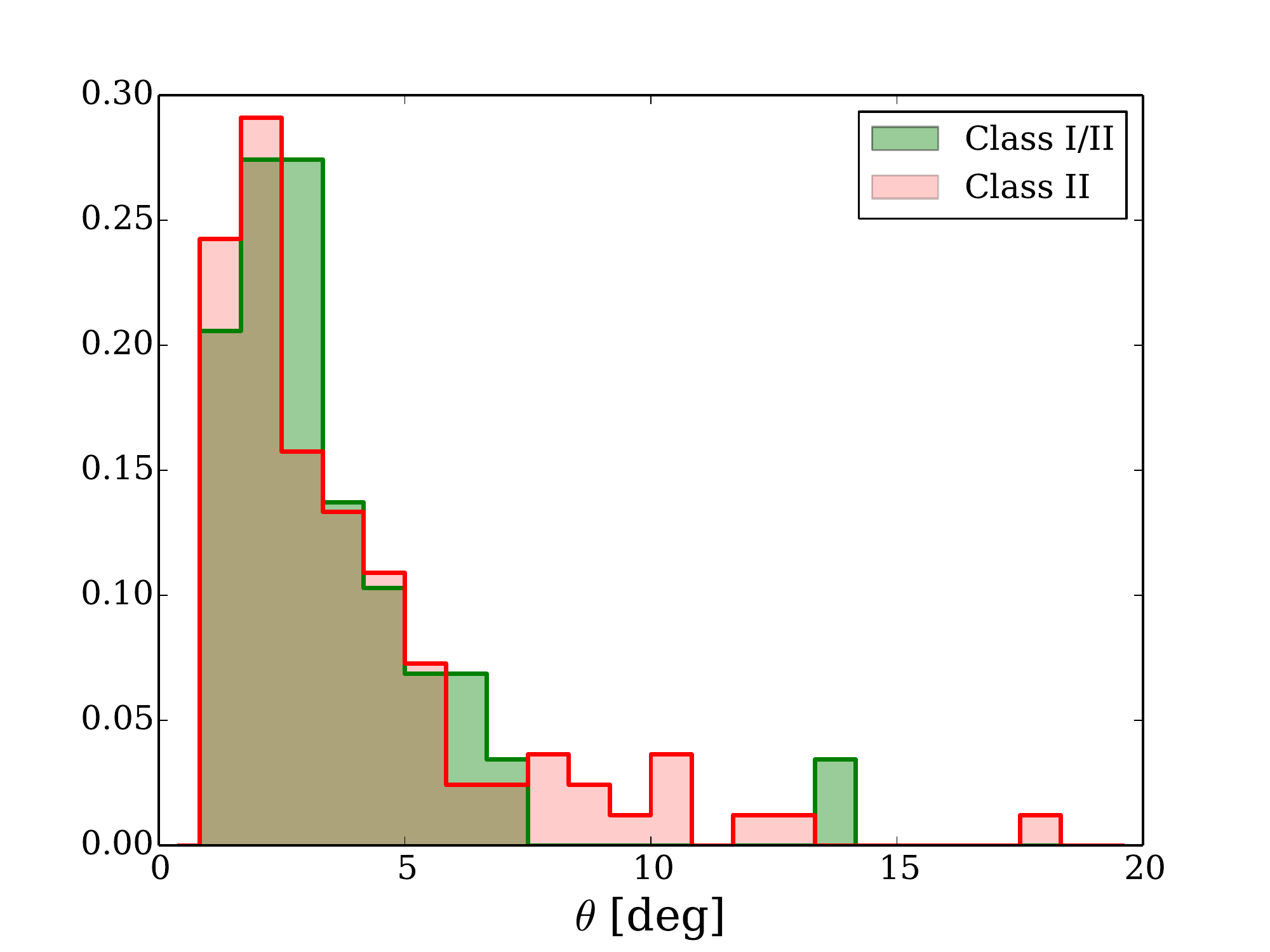}
\caption{Distribution of the most probable angles $\theta$ for classes I/II and II. Areas are normalised.}
\label{fig::distribution_angles}
\end{figure}

This method is very effective for knots with high apparent velocities with a well-peaked probability at low angles. This efficiency decreases at low velocities and, because of the weak beaming constraint, the probability tends to the isotropic distribution. 
The angular distribution of class I/II and II sources deduced from this method is represented in Figure \ref{fig::distribution_angles}.

\begin{figure}[h!]
      \centering \includegraphics[width=9cm]{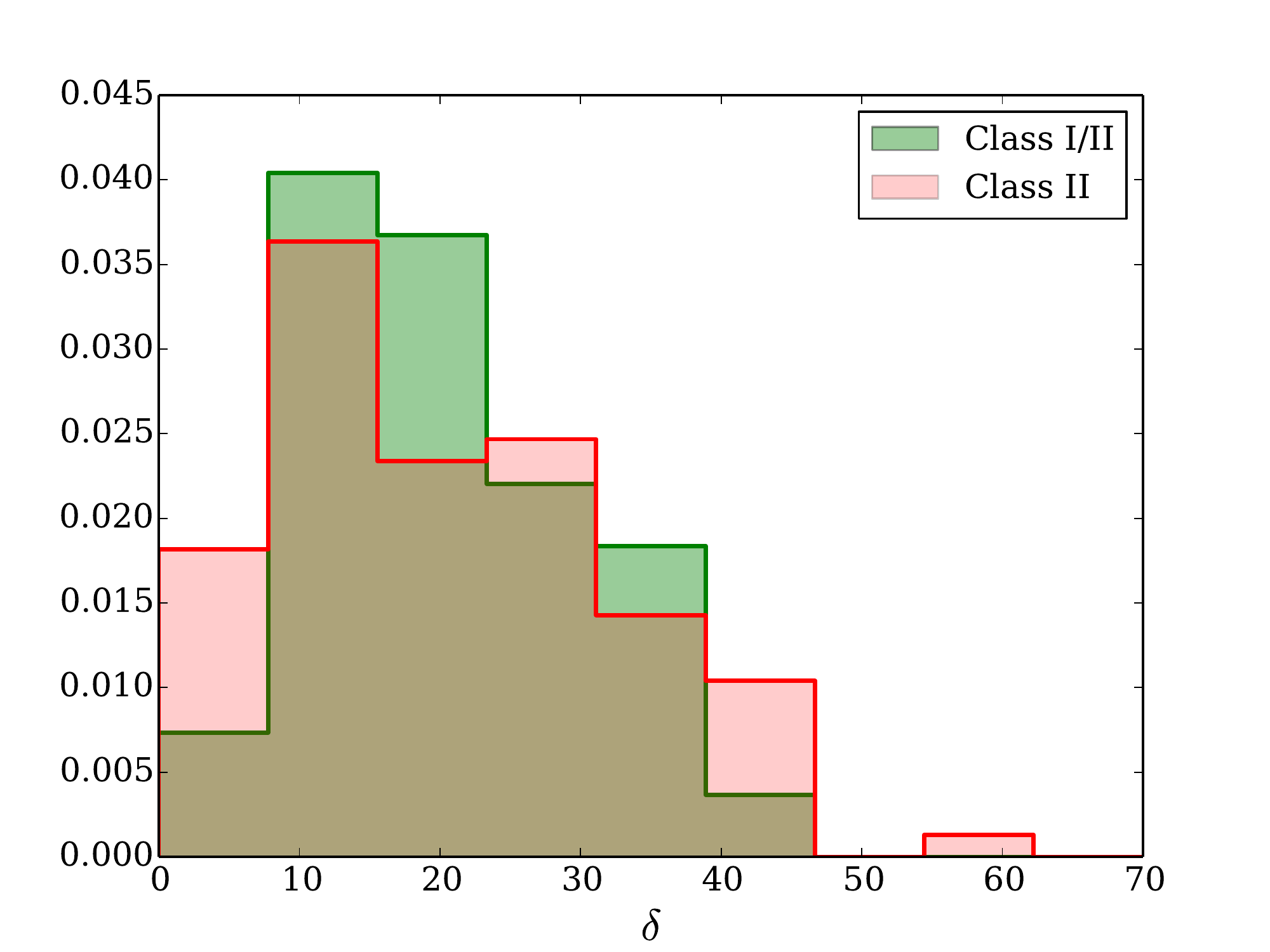}  
      \centering \includegraphics[width=9cm]{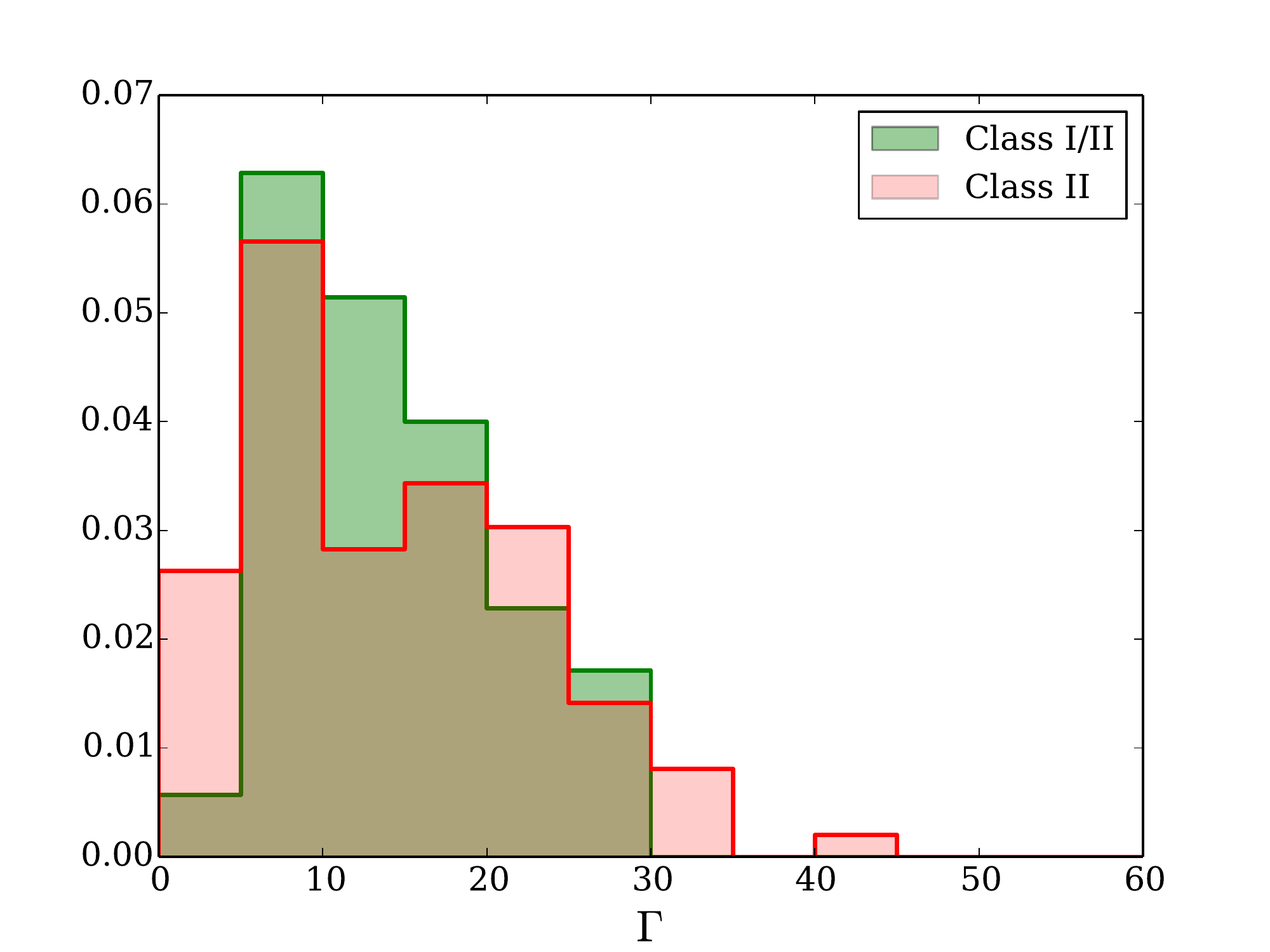}
      \caption{Distribution of the most probable Doppler and Lorentz factors for classes I/II and II. Areas are normalised.}
   \label{fig::distribution_lorentz}
\end{figure}

 We can deduce the Doppler factor distribution from this angular distribution via Eq. \ref{eq::dopp_bapp}. Similarly, we also deduce the Lorentz factor distribution via the equation
\begin{equation}
\Gamma(\theta,\beta_{app}) = \frac{1}{\sqrt{1-\left(\dfrac{\sin \theta}{\beta_{app}} + \cos \theta \right)^{-2}}}.
\label{eq::Lorentz_theta_beta}
\end{equation}

Doppler and Lorentz factors distributions are shown in Figure \ref{fig::distribution_lorentz}, which shows that the distribution peaks of class I/II and II coincide, suggesting similar launching processes of the flow.  Class II presents slightly wider distributions than those of class I/II.

\begin{figure}[h]
\centering \includegraphics[width=9cm]{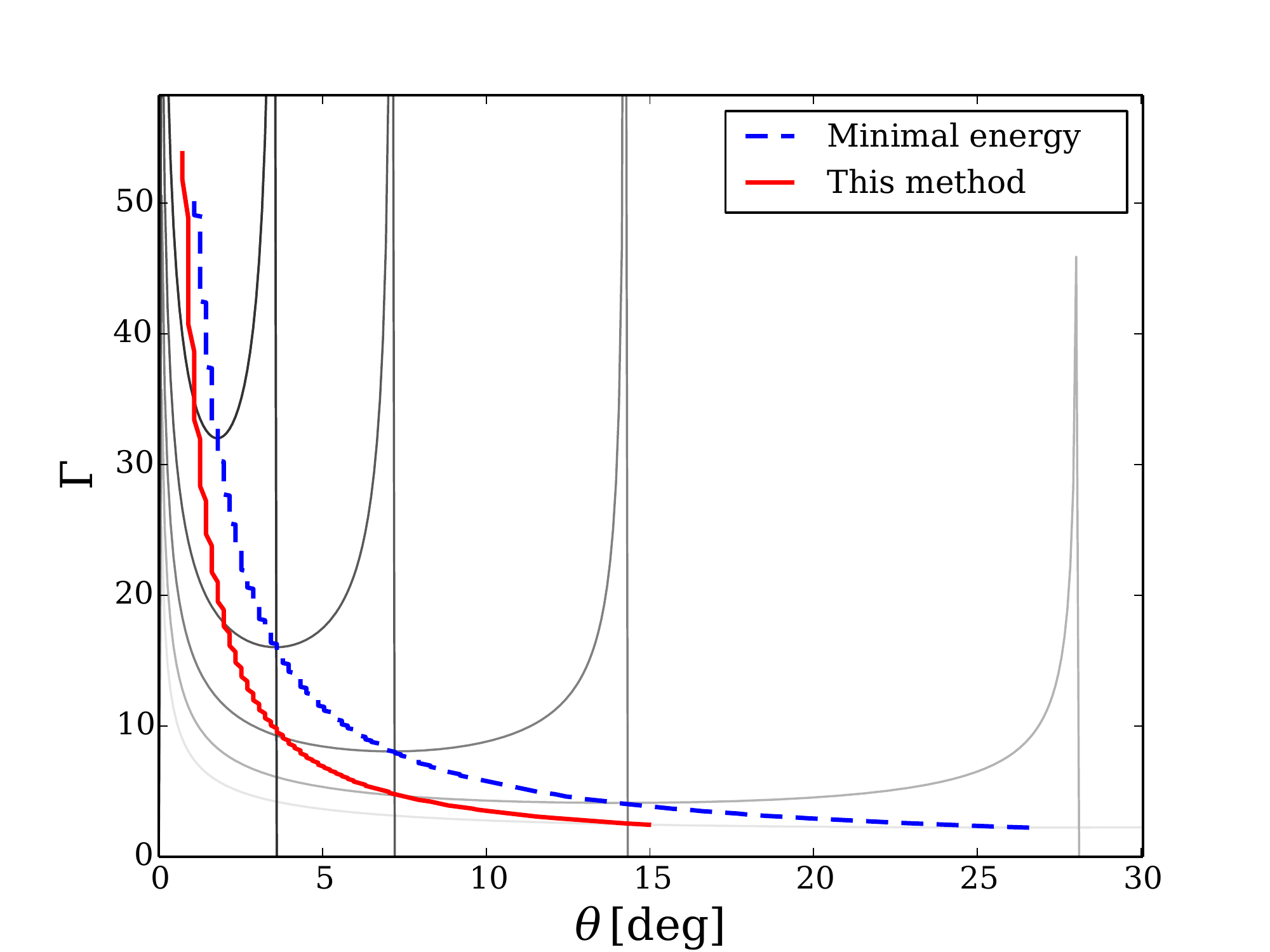}
\caption{Lorentz factor and jet angle for various values of apparent velocities $\beta_{app} \in [2 ; 50]$. We compare results from our method developed here and the minimal energy method. Greyscale lines represent the Lorentz factor values following jet angles for a sample of given apparent velocities (lower to upper lines: 2 c, 4 c, 8 c, 16 c, and 32 c).}
\label{fig::lorentz_vs_theta_Emin}
\end{figure}

Our method can be compared to the method that  minimises the intrinsic kinetic power of the jet. From the principle of least action, we can assume that the observation angle is close to the observation angle that minimises the Lorentz factor value for a given apparent velocity.
We present in Figure \ref{fig::lorentz_vs_theta_Emin} the Lorentz factor values for the "minimal energy" and our method. In any case the Lorentz factor found in our method remains close to the lowest possible value, which reinforces the relevance of our statistical approach.
All of the observation angles, Doppler and Lorentz factor deduced in this section are reported in Table \ref{Table::bigtable}.


\subsection{Inner jets aperture angles}
 
In \cite{Hervet_2015}, we showed that the radio knots of the blazar AP Librae increases linearly with the radio core distance. This allowed us to define an aperture angle of the inner jet. Now we apply this method for the blazars of our three classes to understand whether these aperture angles are dependent on the knot kinematics.

The radio knot flux distributions on the sky plane are fitted by 2D symmetrical Gaussians \citep{Lister_2013}, thus we can estimate the apparent half aperture angles by making a linear regression of their half full-width at half-maximum (FWHM) following their distances to the radio core. This method was already used in \cite{Jorstad_2005} and \cite{Hervet_2015}.

\begin{figure}[h]
\includegraphics[width=9.0cm]{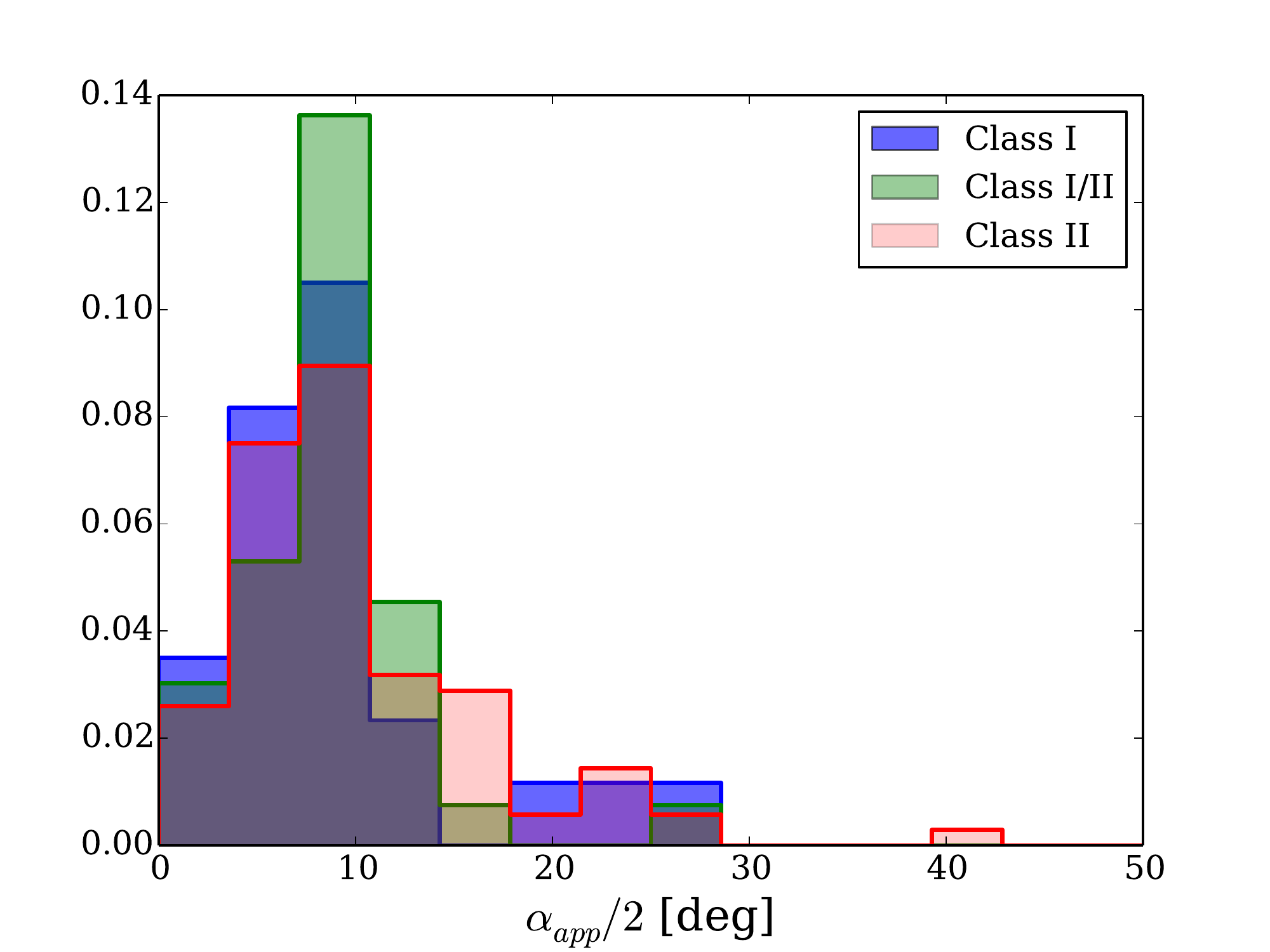}
\caption{Inner jet half opening angles distribution for classes I, I/II et II. Areas are normalised.}
\label{Fig::rep_ouv_app}
\end{figure}

Sources with a correlation coefficient of the linear regression, $R^2 > 0.1$, are selected in our sample, and only three sources do not pass this selection; thus, this gives us  very strong confidence in the conical description of blazar jets.
The apparent angles distribution deduced from this method is shown for each class in Figure \ref{Fig::rep_ouv_app}, and reported source by source in Table \ref{Table::bigtable}.

\begin{figure}[h]
\includegraphics[width=9.0cm]{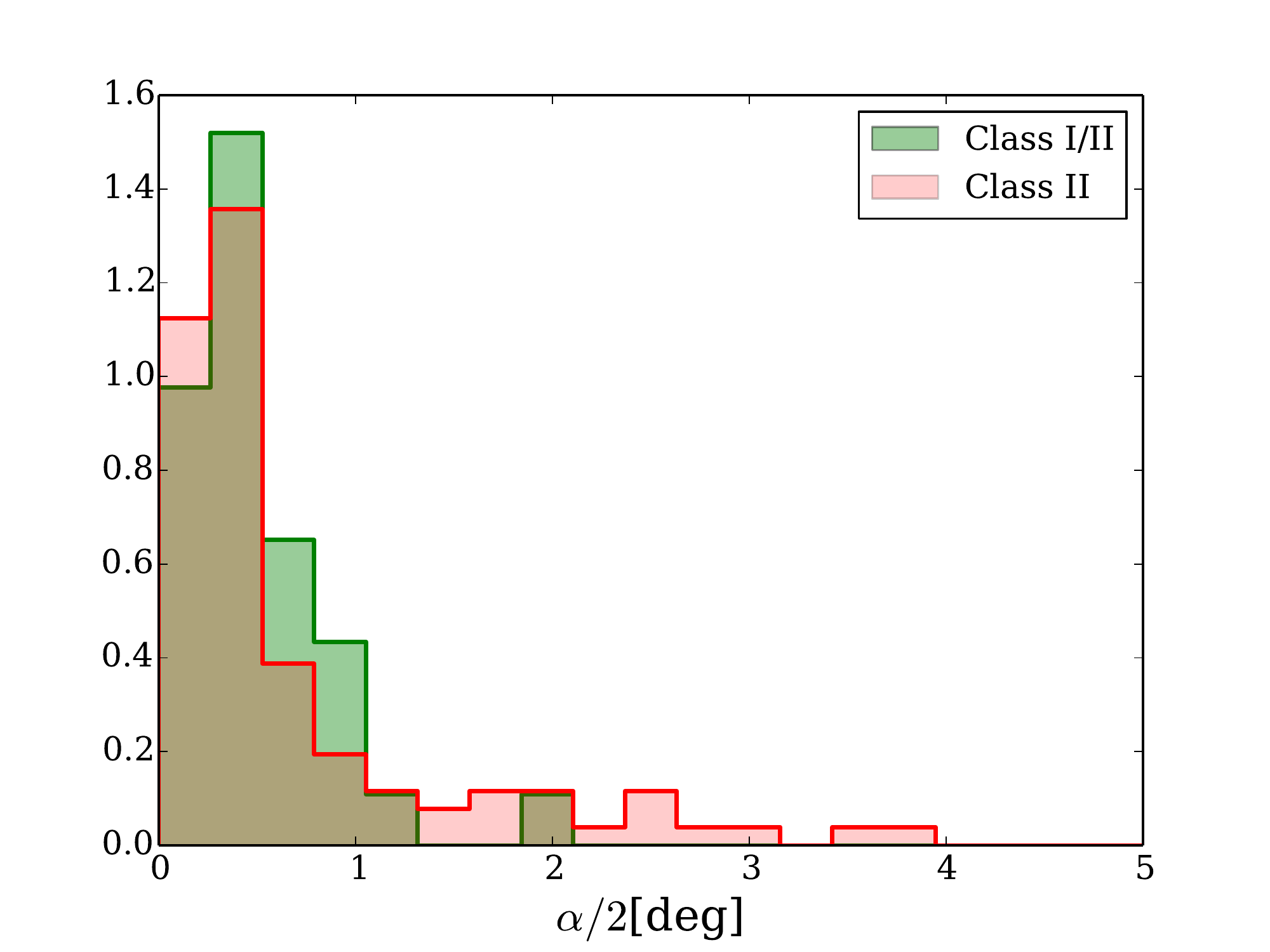}
\caption{Intrinsic half opening angles distribution for classes I/II et II. Areas are normalised.}
\label{Fig::rep_ouv}
\end{figure}

Knowing the apparent aperture angle $\alpha_{app}$ and the jet angle to the line-of-sight $\theta,$ calculated in Section \ref{Section::delta_Gamma_alpha}, we can deduce the intrinsic aperture angle of the inner jet $\alpha = \alpha_{app} sin(\theta)$. Having no strong constraints on the $\theta$ angle for class I sources, we present these intrinsic angles for classes I/II and II only.
The distribution of $\alpha$ angles is shown in Figure \ref{Fig::rep_ouv} and the values are given in Table \ref{Table::bigtable}.

\subsection{Radio knot evolution}
\label{Section::Radio knots evolution}

In previous sections we showed that classes I/II and II do not present any significant differences in terms of jet angles with the line of sight, Doppler factors, Lorentz factors, or inner jet aperture angles.
Thus, the kinematical differences between these classes cannot be strongly linked with any of these parameters.

It is also possible to study the knot evolution for the three classes with the radio knot features given in \cite{Lister_2013}, such as radio flux and Gaussian FWHM, which we can link to their sizes.
The knot-core distance evolution is not used in this study because of its strong dependence on the observation angles, poorly constrained in class I sources.
Hence, we determine  the flux and size median evolutions,  $dF_k/dt$ and $dD_k/dt$, for each source, respectively. Their importance is balanced by their visibility time in radio to avoid an over-representation
of some knots. Of course, only knots identified during multiple observations are taken into account.



\begin{figure}[h]
\centering \includegraphics[width=9cm]{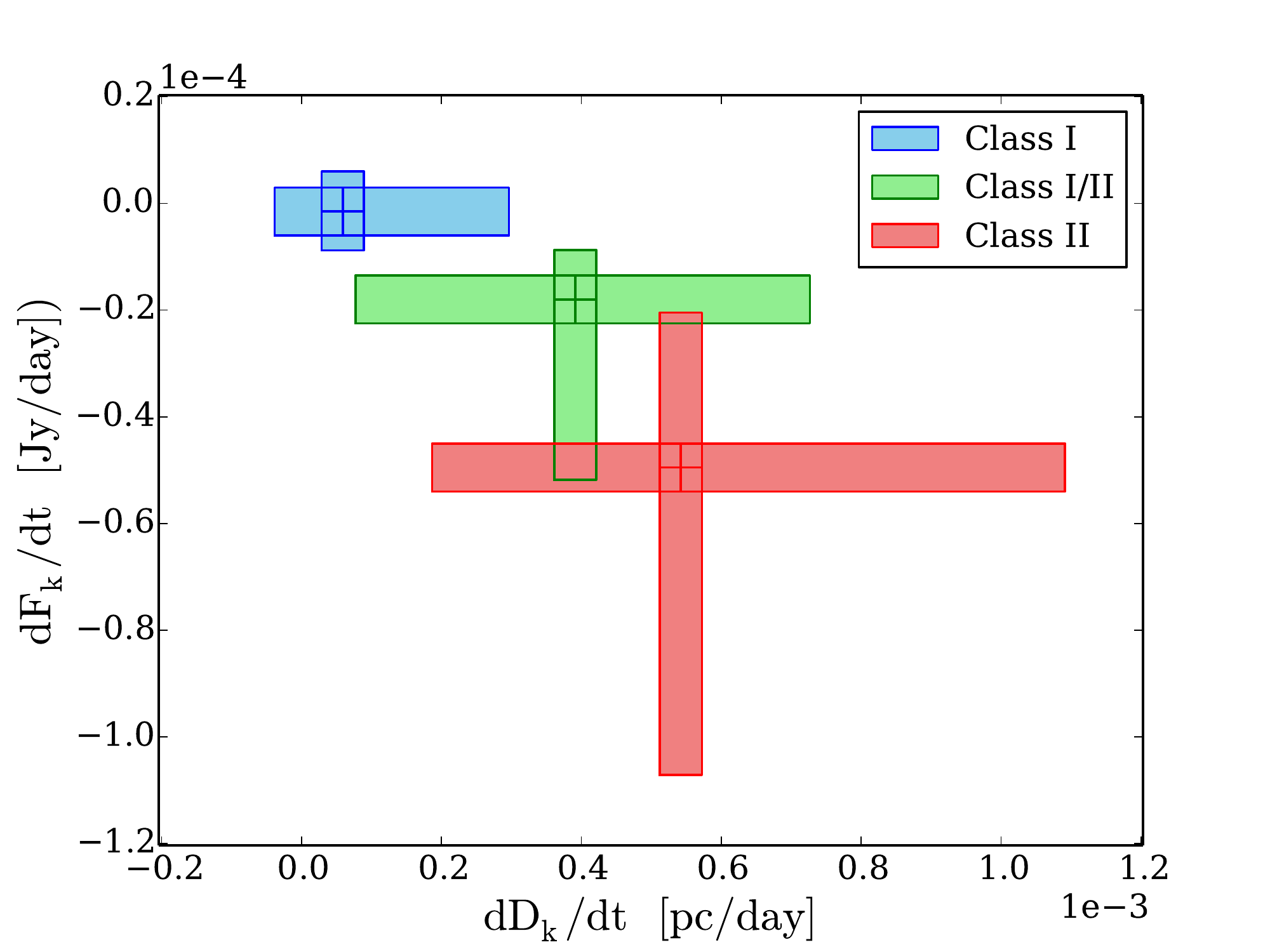}
\caption{Box plots of knot diameter evolution vs. knot flux evolution for sources of the three kinematic classes. Boxes are delimited by
the first and last quartiles; middle lines are medians.}
\label{fig::evolution_moustache}
\end{figure}

Figure \ref{fig::evolution_moustache} shows a clear distinction between the three kinematic classes. The median values of class I sources show an almost non-evolution of their knots, regarding their flux or size.  
The other classes show a clear expansion and cooling with values increasing continuously between class I/II and II.
 Median values of size and flux evolutions are referenced in Table \ref{table::evolution}, quantifying the continuity between dominant regimes for the three classes.

\begin{table}[h]
\centering
\begin{tabular}{ccc}
\hline\hline
\noalign{\smallskip}
Class & $dF_k/dt$ [Jy/day] & $dD_k/dt$ [pc/day] \\
\hline
\noalign{\smallskip}
I & $-1.52\times 10^{-6}$ & $5.89\times 10^{-5}$ \\
II & $-4.95\times 10^{-5}$ & $5.42\times 10^{-4}$ \\
I/II & $-1.80\times 10^{-5}$ & $3.91\times 10^{-4}$\\
\hline
\end{tabular}
\caption{Median values of the size and flux evolution of radio knots for the three classes, illustrating a continuity between VLBI jet behaviours.}
\label{table::evolution}
\end{table}

This confirms that knots with quasi-stationary motion at the jet base also have a quasi-stationary evolution without cooling nor expansion contrary to knots expelled in jets of classes I/II and II, which mainly follow an adiabatic expansion. This implies that knots of class I sources have a stable energy tank over long periods. The description of these knots as structural stable stationary shocks in jets powered by a continuous underlying flow, as developed in the following sections, happens to be the most likely explanation of this phenomenon.

\section{Study of an intermediate blazar, the BL Lac case}
\label{Section::Study of an intermediate blazar, the BL Lac case}

The hybrid kinematics of class I/II VLBI jets is very surprising. Intuitively one would expect a smooth kinematical transition between classes I and II with knots at intermediate velocities, but this is not the case and Lorentz factor values of class I/II, which were deduced in Section \ref{Section::delta_Gamma_alpha}, are not significantly different from those of class II.
This hybrid behaviour with quasi-stationary knots near the base and relativistic knots downstream seems to give a valuable clue regarding the intrinsic nature of these zones. 

To further study these objects, we now focus on the blazar BL Lac itself, which is the most studied blazar of the intermediate class, and presents one of the best radio VLBI monitoring over numerous years.

\begin{figure}[h]
\centering \includegraphics[width=9cm]{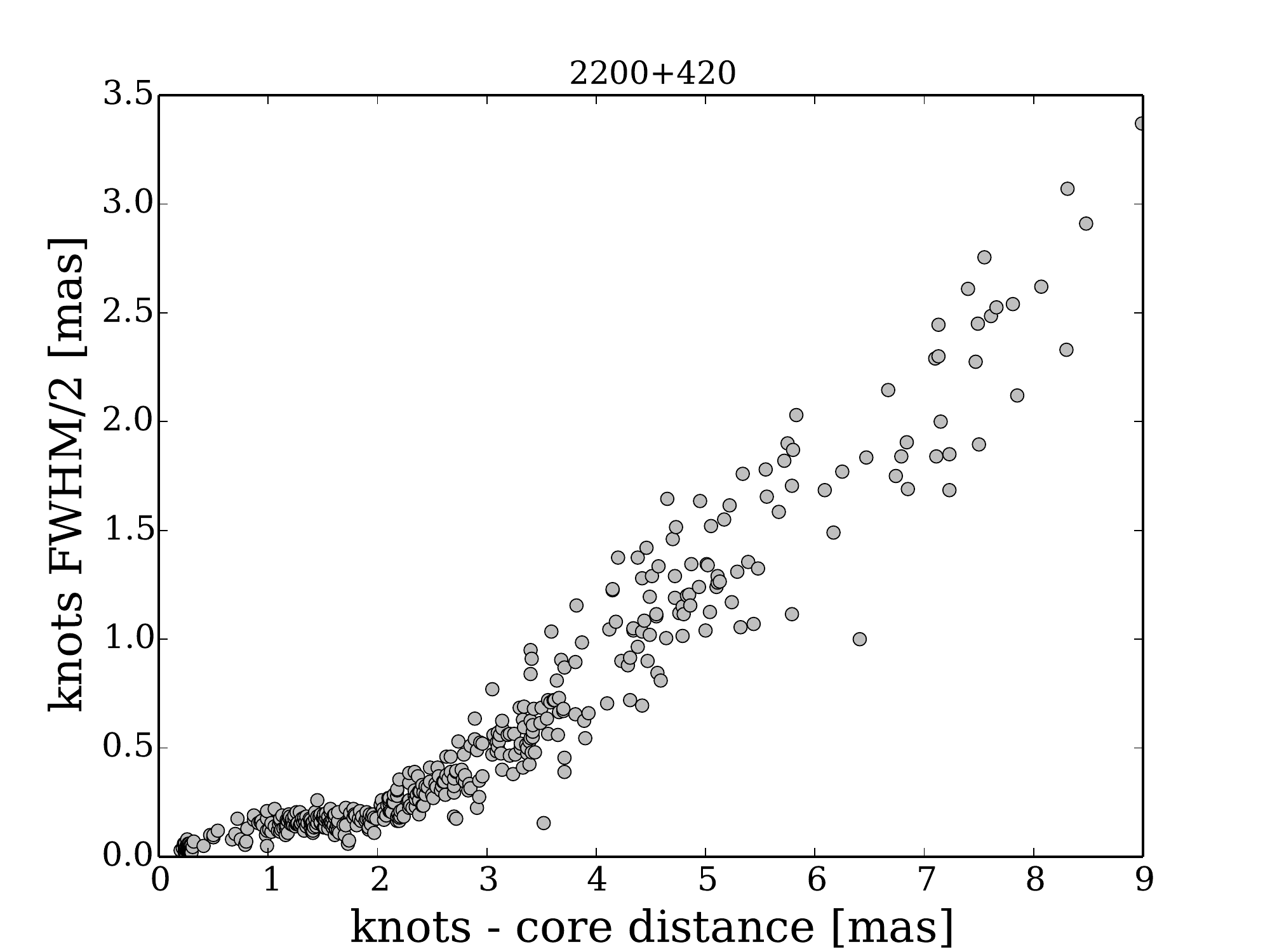}
\caption{Half FWHMs of BL Lac radio knots following their apparent distance to the radio core. A clear change of the opening angle happens around 2 mas from the core. Adapted from data published in \cite{Lister_2013}.}
\label{fig::jet_interne_bl_lac}
\end{figure}

In Figure \ref{fig::jet_interne_bl_lac} we notice a strong increase of the inner jet aperture angle for the apparent core distance $L_{app,jet} = 2$ mas. By linear approximation, we deduce two apparent aperture angles $\alpha_{app,1}/2 = 4.6^\circ$ and $\alpha_{app,2}/2 = 21.5^\circ$ upstream and downstream 2 mas.

\subsection{Transient knots and propagation of a perturbation }

\begin{figure*}[!b]
\centering \includegraphics[width= 16cm,trim= 4cm 0cm 0cm 0cm, clip]{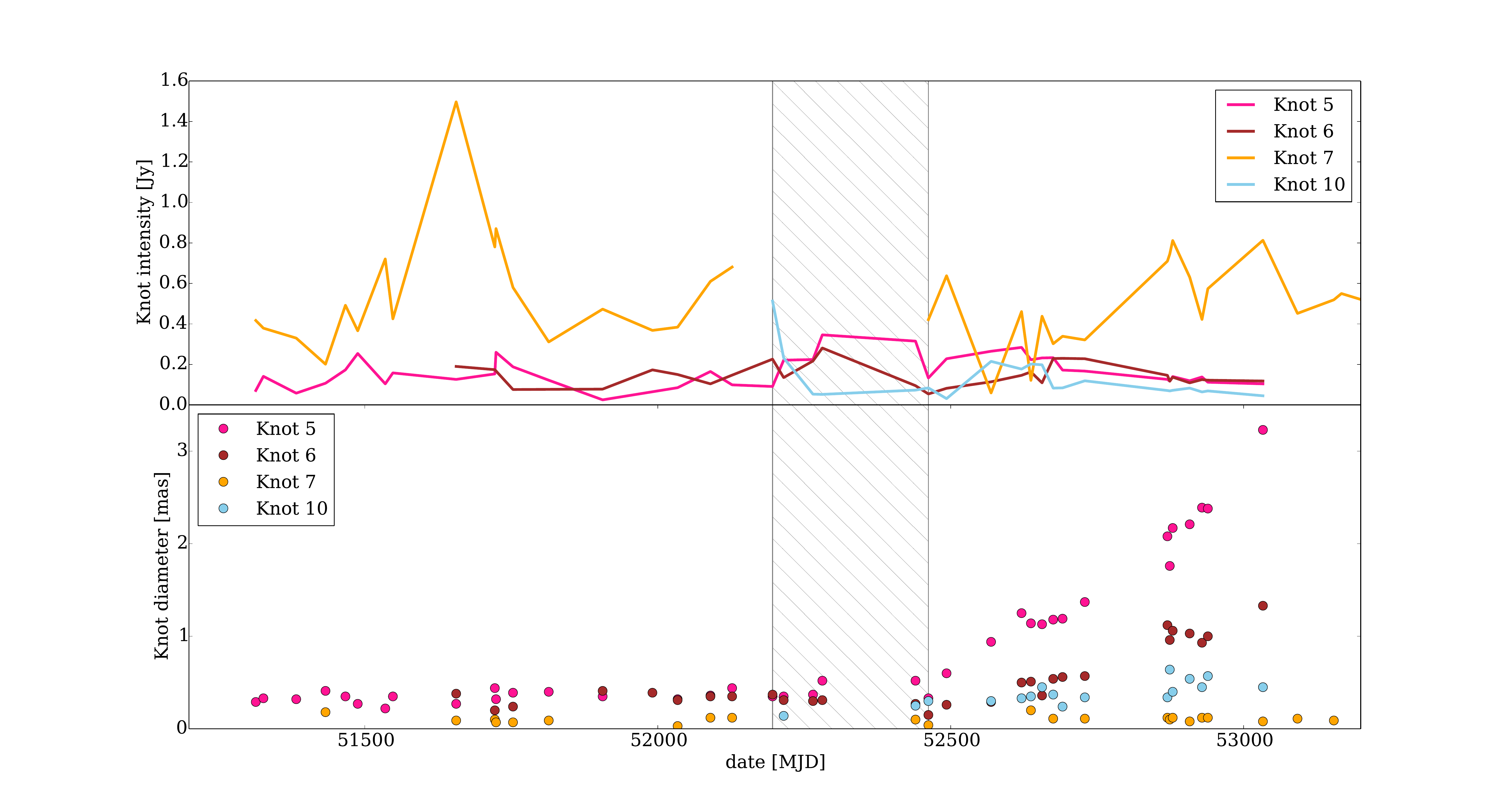}
\caption{Evolution of flux and size of four radio knots of BL Lac 5,6,7 and 10. The grey hatching area represents the duration of the perturbation propagation in the VLBI jet from the knot 10 to the knot 5 between the October 14th 2001 and the July 7th 2002. The non-observation of the knot 7 and the appearance of the knot 10 during the perturbation propagation is a sign of an abrupt evolution of the same knot.}
\label{fig::bllac_transient_knots}
\end{figure*}

\begin{figure}[h]
\centering \includegraphics[width=7cm]{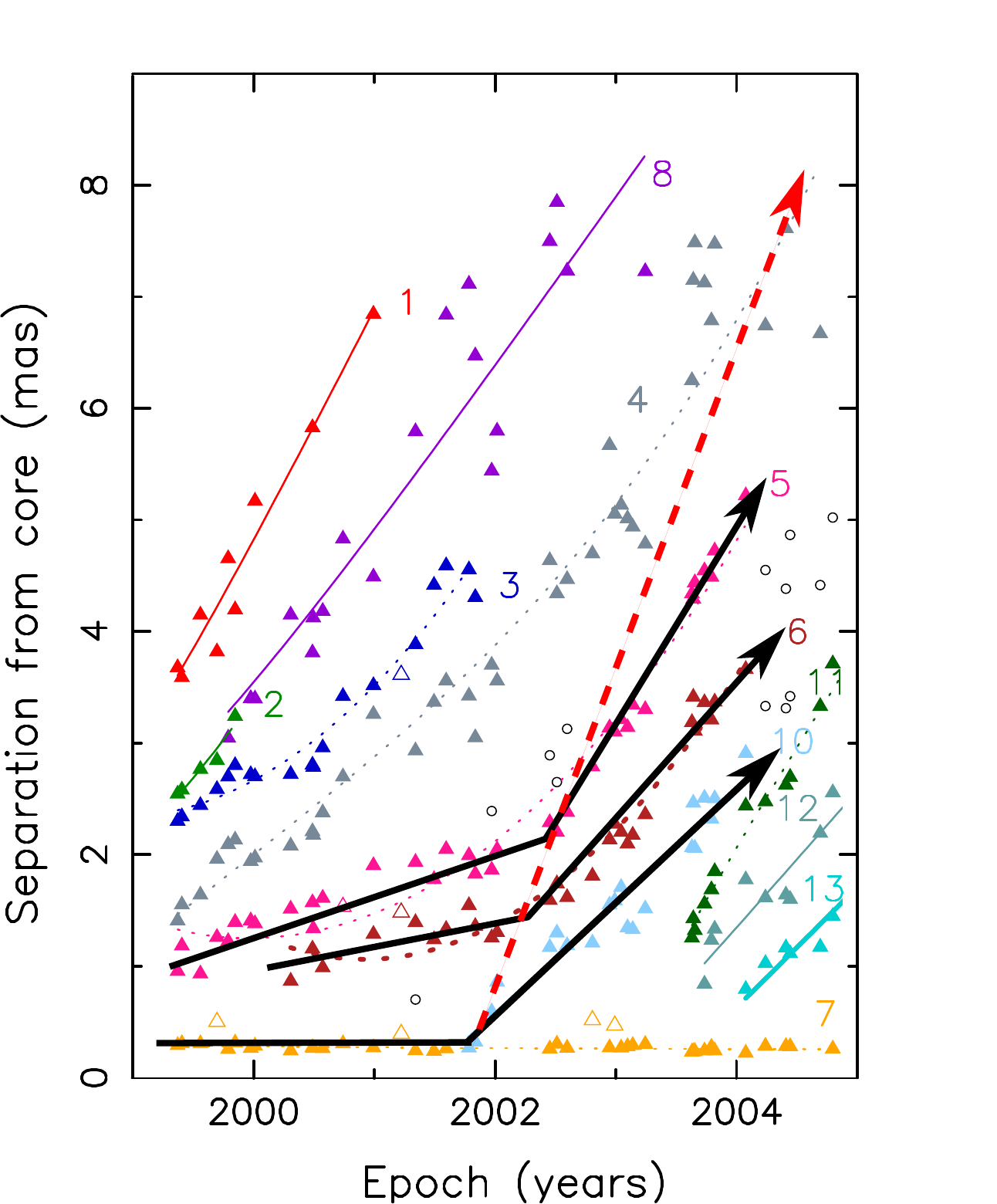}
\caption{Knot distances to the radio core of BL Lac over 7 years, adapted from \cite{Lister_2013}. Knots 5 and 6 are identified as transients; knot 10 seems ejected from the closest stationary zone from the core. The dashed red arrow represents the displacement of a probable perturbation along the jet having modified the kinematics of knots.}
\label{fig::cinetique_bllac}
\end{figure}

Some of the apparent motions of BL Lac radio knots  present a quick evolution. Indeed, some long-term quasi-stationary knots are quickly accelerated to relativistic velocities, as shown in Figure \ref{fig::cinetique_bllac}. We call these "transient knots" because they exhibit a transition between knots of class I and those of the class II.

Figure \ref{fig::cinetique_bllac} shows that three knots (5,6 and 10) are accelerated in a short time around the beginning of 2002. This suggests that the internal jet undergoes a perturbation on several parsecs during few months in the observer frame.
We estimate a temporal gap of 266 days between the ejection of knot 10 at the base of the jet on October 14, 2001 and the acceleration of knot 6 at 2.6 mas from the core around July 7, 2002. The apparent velocity of such a propagation would be $\beta_{app,P} = 14.9$, which is significantly higher than the highest apparent velocity measured in VLBI at 9.95 c.

The propagation velocity of a perturbation could be a more reliable indicator of the intrinsic flow velocity than the knot motions. Indeed the radio knots are supposed to have particle density that is higher than the underlying flow. Admitting that this flow carries the knots along, the radio knot velocities should systematically underestimate the flow velocity.
As presented in Section \ref{Section::Relevance of a kinematical blazar classification}, the fastest radio knot is chosen to minimise this bias.

This effect can be balanced, however, especially for class II sources. In these jets, radio knots have relativistic velocities from the radio core and propagate over large distances, and we can thus assume that they reach  a kinetic equilibrium with the underlying flow more quickly and, thereby, are more effective velocity markers. 
With this direct indicator of the intrinsic flow velocity in the blazar BL Lac, we can constrain various physical parameters, as developed in Section \ref{Section::delta_Gamma_alpha} (see Table \ref{table::physique_Bllac}).

\begin{table}[h]
\centering
\caption{Physical parameters of BL Lac deduced from the fast propagation of a perturbation in the jet. $\alpha_1/2$ and $\alpha_2/2$ are the intrinsic half aperture angles of the inner jet for apparent distance to the core  upstream and downstream 2 mas, respectively.}
\label{table::physique_Bllac}
\begin{tabular}{ccc}
\hline\hline
\noalign{\smallskip}
Parameter & Value & Unit \\
\hline
\noalign{\smallskip}
$\theta$ & $2.2$ & deg \\
$\delta$ & $23.8$ & \\
$\Gamma$ & $16.6$ &  \\
$\alpha_1/2$ & $0.17$ & deg \\
$\alpha_2/2$ & $0.81$ & deg \\
\hline
\end{tabular}
\end{table}

\subsection{Transient knot evolution}

The effects of the perturbation described above on the knot evolution is now studied. The intrinsic properties of radio knots can be defined by two observables, size and flux evolution; these are presented in Figure \ref{fig::bllac_transient_knots}.

We notice that knot 10 shows a high flux when detected and then decreases quickly. During its ejection (0.513 mJy and 0.14 mas, respectively), the flux and size of knot 10 are relatively close to those of the stationary knot 7 just before (0.681 mJy et 0.12 mas).
Also, knot 7 is no longer detected during the four following observations between  August 6, 2001 and June 15, 2002.
Thus, all these evidences suggest that knot 10 is nothing else than knot 7 ejected from its stationary zone accompanied by a strong decrease in flux. 
The effect of this perturbation on knots 5 and 6 farther from the core is less obvious with a slight increase and decrease of their flux during the interval estimated for the propagation of this perturbation.

One can also see in Figure \ref{fig::bllac_transient_knots} that the three knots (5, 6, and 10) are in expansion regime after the perturbation passage. This is consistent with their motion observed in the jet. Knot 5 shows a faster expansion in accordance with the increase of the inner jet opening angle after 2 mas (see Figure \ref{fig::jet_interne_bl_lac}).
The reobservation of stationary knot 7 after the perturbation indicates that this unstable zone is naturally reforming, highlighting an intrinsic structure of the jet, which we  develop in the following section.

\section{Discussion and interpretation}
\label{Section::Discussion and interpretation}

We propose here a qualitative scenario that is able to take the various radio knot characteristics discussed above into account. 
Observed apparent velocities, variabilities, and also multi-wavelength modellings prove that AGN jets host strong flows with relativistic velocities.
As these velocities are usually much higher than the Alfven velocity, it is very likely to have the formation of recollimation shocks.
Such shocks are commonly invoked for the radio core description \citep{Marscher_1985} or for some stationary knots as HST1 in M87 \citep{Bromberg_2009} and knot 7 of BL Lac \citep[C7 in][]{Cohen_2014}.
These scenarios are based on the presence of one powerful recollimation shock that is able to change the jet structure; this is usually known as the master recollimation shock (MRS). The stationary knot string structure present in class I and I/II sources, as defined in this paper, provides evidence about the possible presence of multiple recollimation shocks in jets.
These multi-shocks structures are also predicted by relativistic MHD simulations. For example, \cite{Mizuno_2015} recently highlighted the impact of magnetic field topologies on the efficiency of recollimation shock on particle acceleration.
These authors show that a longitudinal magnetic field induces more powerful recollimation shocks than any other magnetic topology. Otherwise, magnetic field topologies deduced from the Faraday rotation in radio jets by \cite{Kharb_2008B,Kharb_2008A} and confirmed by \cite{Gabuzda_2014} indicate that HBLs are dominated by a longitudinal field. In contrast, LBLs and FSRQs present a two-component structure with an inner jet that is dominated by a toroidal field included in a wider external jet that is dominated by a longitudinal field.

These various observed magnetic topologies support the scenario of multiple recollimation shocks. 
Dominant longitudinal fields observed in HBLs are more able to effectively accelerate particles in shocks, and multiple shock structures, as in class I sources, can further increase the particle acceleration \citep{Meli_2013}.
The VHE properties of HBLs suggest that these objects are, by nature, the most efficient particle accelerators of all of the blazars, which is consistent with that scheme.
Thus, the synchrotron peak frequency appears to be linked to the shock efficiency, depending on the jet magnetic topologies.

\begin{figure}[h]
\centering \includegraphics[width=9cm]{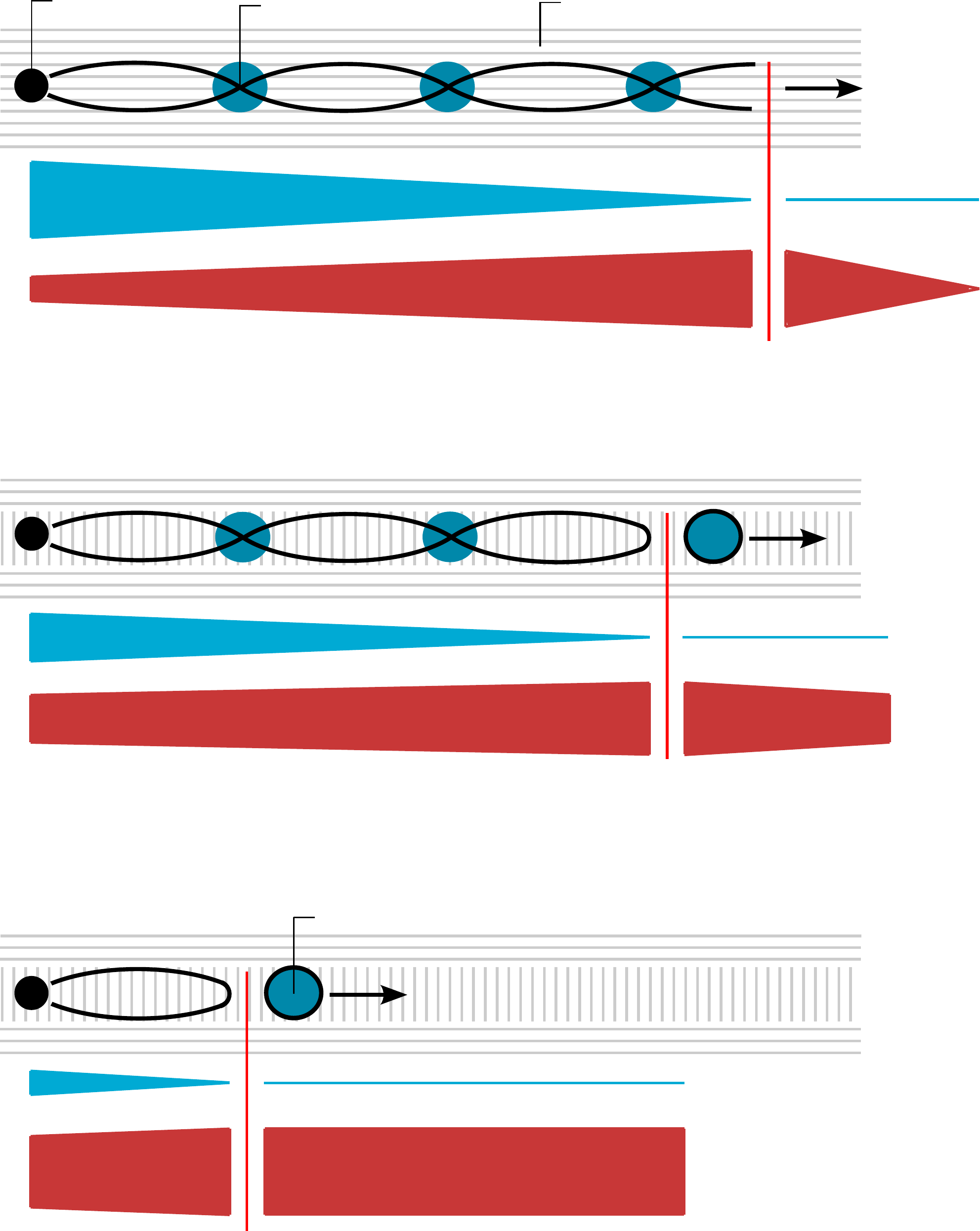}
\put(-160,335){\color[rgb]{0,0,0}\makebox(0,0)[lb]{\smash{\textbf{Class I}}}}
\put(-263,268){\color[rgb]{0,0,1}\makebox(0,0)[lb]{\smash{$\mathbf{E_m}$}}}
\put(-263,246){\color[rgb]{0.5,0,0}\makebox(0,0)[lb]{\smash{$\mathbf{E_c}$}}}
\put(-160,210){\color[rgb]{0,0,0}\makebox(0,0)[lb]{\smash{\textbf{Class I/II}}}}
\put(-263,154){\color[rgb]{0,0,1}\makebox(0,0)[lb]{\smash{$\mathbf{E_m}$}}}
\put(-263,132){\color[rgb]{0.5,0,0}\makebox(0,0)[lb]{\smash{$\mathbf{E_c}$}}}
\put(-160,98){\color[rgb]{0,0,0}\makebox(0,0)[lb]{\smash{\textbf{Class II}}}}
\put(-263,36){\color[rgb]{0,0,1}\makebox(0,0)[lb]{\smash{$\mathbf{E_m}$}}}
\put(-263,13){\color[rgb]{0.5,0,0}\makebox(0,0)[lb]{\smash{$\mathbf{E_c}$}}}
\put(-242,318){\color[rgb]{0,0,0}\makebox(0,0)[lb]{\smash{SMBH}}}
\put(-187,318){\color[rgb]{0,0,0}\makebox(0,0)[lb]{\smash{stationary knot}}}
\put(-108,318){\color[rgb]{0,0,0}\makebox(0,0)[lb]{\smash{magnetic topology}}}
\put(-170,81){\color[rgb]{0,0,0}\makebox(0,0)[lb]{\smash{moving knot}}}
\caption{Scheme of the three kinematic classes I, I/II, and II. We suppose that the various kinematics come from various balances between the internal jet magnetic energy, $E_m$, and the kinetic energy, $E_c$. The width of these components is representative to their relative strength along the jets.}
\label{fig::schema_recollimation}
\end{figure}
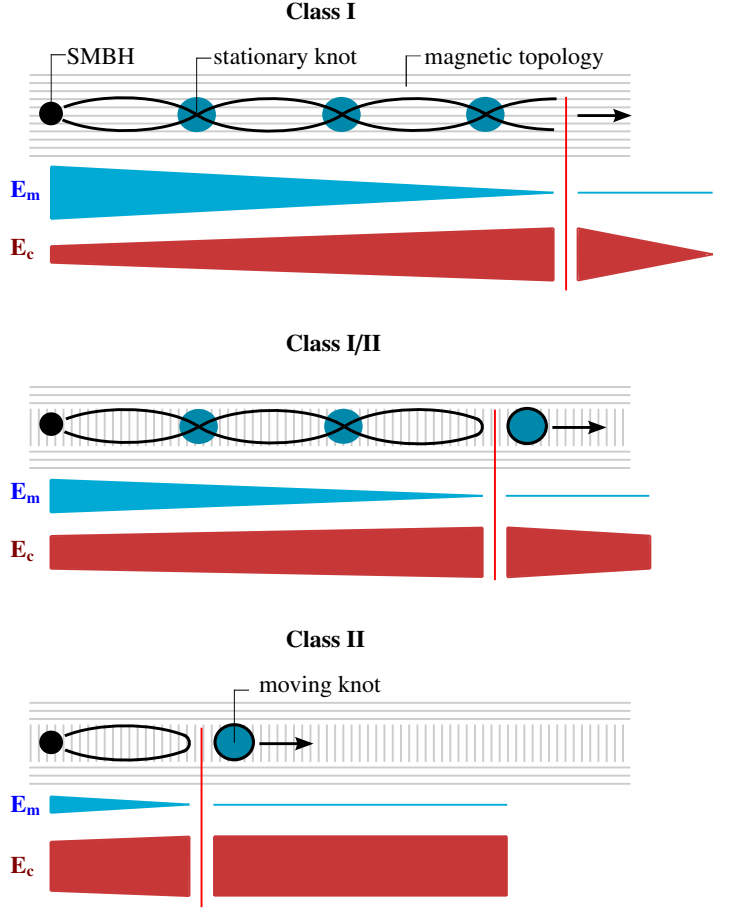

We focus now on various kinematics of radio knots. A key point in differentiating blazars is the jet power. It is recognised that spectral classes are a function of the jet power, that is $P_{FSRQ} > P_{LBL/IBL} > P_{HBL}$ \citep{Meyer_2011,Celotti_2008}. However, these powers alone do not explain why some knots are stationary while others are not. The kinematic classification presented in this paper supports an interpretation as described below. We give a schematic view of this scheme in Figure \ref{fig::schema_recollimation}.
This scheme does not intend to describe the complexity of each individual source of this study, but summarises the dominating behaviour of the three blazar populations considered for this classification.

\begin{itemize}

\item[$\bullet$] \textbf{Kinematic class I:} The presence of stable recollimation shocks in class I sources, as shown in Section \ref{Section::Radio knots evolution}, suggests that the longitudinal magnetic field plays a more important role than in other sources. A strong magnetic compression induced by a longitudinal field associated with a low kinetic power maintains a stable structure along the jet up to the magnetic field dissipation after successive shocks, in which particle acceleration occurs at the expense of the magnetic energy. In accordance to this
interpretation, HBL sources that belong to class I are known to be closer to the equipartition than FSRQs with a more significant magnetic field.
Also, the peculiar large-scale low luminosity of class I, shown in Figure \ref{fig::R_vs_l_ext}, is in agreement with this peculiar magnetic topology, which should induce a strong dissipation of energy before the large scales.

\item[$\bullet$] \textbf{Kinematic class II: } Associated with class II, FSRQs have jets that are widely dominated by particle kinetic energy and do not allow the existence of a successive standing shock structure. Recollimation shocks can be generated at the jet base, where the magnetic field is the most powerful, but seems highly unstable and systematically carried downstream by the underlying flow.
As these shock zones are no longer magnetically constrained, they follow an adiabatic expansion along jets, as shown in Section \ref{Section::Radio knots evolution}.

\item[$\bullet$] \textbf{Intermediate kinematic class I/II:} Intermediate sources are very interesting because they have stationary knots at the jet base and relativistic velocities detected downstream. We assume that the magnetic energy is sufficient near the jet base to maintain a stationary shock structure.
This energy is however dissipated in each shock into radiative and kinetic energy. From a certain distance to the core, the magnetic structure becomes unstable and shocks are finally carried out in the same way as those of class II before they are naturally reformed, since they depend on the internal structure and power equilibrium of jets.
\end{itemize}

Another aspect to take into account is that sources of classes I/II and II are more likely to have an imbricated jet structure. This structure generates instabilities because the shearing is able to break the recollimation shock strings of the internal jet, but maintain a good general collimation because of its strong kinetical power \citep{Meliani_2007,Meliani_2009}.

This general interpretation, which is consistent with the observations, brings the idea that blazars jets are not described  by one powerful recollimation shock alone, but the number and kinematics of these recollimation shocks depends on the energetic balances of jets. 
Furthermore, a newly published paper from \cite{Gomez_2016}, regarding the ultra high resolution of the BL Lac jet, reports on the detection of two other stationary knots that are closer to the core than knot 7 described above; these authors associate knot 7 with several recollimation shocks. This study reinforces our proposed scenario describing  class I/II.

\section{Conclusion}

The usual blazar classification based on the SED shape informs the microphysics of a small portion at high energy of the total particle population of jets.
Conversely, the new kinematical classification proposed in this paper, provides macrophysics information about the structure and propagation of jets.
Because the VLBI resolution is able to access projected sizes at the submilliarcsecond scale, we can reach the regions closest to those responsible for high energy emissions.

The link that we found between spectral and kinematic classifications (class I corresponding to HBLs, class I/II corresponding to IBLs and LBLs, and class II corresponding to FSRQs) gives valuable clues to the area and mechanisms responsible for the particle acceleration. A scenario of multiple recollimation shocks is favoured to interpret this link and also to describe the various behaviours of VLBI jets. We highlight some long-standing issues of AGN classification, which are naturally explained within such a scenario:

\begin{itemize}
\item[$\bullet$] It can resolve the apparent Lorentz factor paradox between jets showing evidence for high Lorentz factor values and quasi-stationary VLBI knots.
\item[$\bullet$] It  account for the singular VLBI behaviour of intermediate sources classified as I/II.
\item[$\bullet$] It can chart the destructuration and restructuration of a knotty VLBI structure after a strong flare, as observed in BL Lac.
\item[$\bullet$] It provides a consistent scheme linking magnetic topology, synchrotron peak frequency, and jet VLBI kinematics.
\end{itemize}

Possible biases that can affect the significance of the results are the  lack of detailed  VLBI monitoring of certain sources that can lead to uncertainties about the detected velocities of the knots. It was shown, however, that these effects are only potentially significant for class I. Thus, the global scenario presented here is not jeopardised.

In any case this scenario  with multiple recollimation shocks in blazars will be tested along several paths.
The MHD modelling aspect will be deepened to test the viability of shocks and to find out, in particular, if we can reasonably reproduce the behaviour of class I/II sources.
Class I knots are associated with multiple stationary shocks. This gives constraints on their sizes and inter-shocks distances, which can be checked by VLBI studies.
Still, in class I sources, one would expect that perturbations propagate in their jets without destructuring them, passing through shocks by increasing their luminosities. Variability measures that are consistent with the time delay between shocks could also efficiently test this scenario.

\begin{acknowledgements}
This research has made use of data from the MOJAVE database that is maintained by the MOJAVE team \citep{Lister_2009}.

\end{acknowledgements}

\bibliographystyle{aa}
\bibliography{Int_blazars}

\onecolumn
\LTcapwidth= 0.8\textwidth
\begin{longtable}{llllccccc}

\caption{\label{Table::bigtable} Kinematic classification and physical parameters of our source sample. Angles with the line-of-sight $\theta$, Doppler factor $\delta$, Lorentz factor $\Gamma$, as well as apparent and intrinsic half opening angles of inner jets $\alpha_{app}/2$ and $\alpha/2$ are calculated in Section \ref{Section::Jets physical parameters}.}\\
\hline\hline
B1950-name & Common name & Spectral & Kinematic & $\theta$ & $\delta$ & $\Gamma$ & $\alpha_{app}/2$  & $\alpha/2$\\
& & class & class & [deg] & & & [deg] & [deg]\\
\hline
\endfirsthead
\hline\hline
B1950-name & Common name & Spectral & Kinematic & $\theta$ & $\delta$ & $\Gamma$ & $\alpha_{app}/2$  & $\alpha/2$\\
& & class & class & [deg] & & & [deg] & [deg]\\
\hline
\endhead
\hline\\
\endfoot
0003-066 & NRAO 005 & LSP BL Lac & I/II & 3.6 & 14.0 & 9.6 & 6.4 & 0.4\\
0003+380 & S4 0003+38  & FSRQ & I/II & 6.5 & 7.8 & 5.3 & 9.9 & 1.13\\
0010+405 & 4C +40.01 & SSRQ & I/II & 4.3 & 11.7 & 7.9 & 2.7 & 0.20\\
0016+731 & S5 0016+73  & FSRQ & II & 3.8 & 13.5 & 9.3 & 28.2 & 1.86\\
0048-097 & PKS 0048-09  & ISP BL Lac & I & - & - & - & 19.5 & -  \\
0059+581 & TXS 0059+581  & FSRQ & II & 3.4 & 14.8 & 10.1 & 9.5 & 0.57\\
0106+013 & 4C +01.02  & FSRQ & II & 1.4 & 36.6 & 26.5 & 11.9 & 0.30\\
0109+224 & S2 0109+22 & LSP BL Lac & I & - & - & - & 8.1 & -\\
0110+318 & 4C +31.03 & SSRQ & II & 1.8 & 28.7 & 20.1 & 3.5 & 0.11\\
0111+021 & UGC 00773  & BL Lac & I & - & - & - & 3.3 & - \\
0119+115 & PKS 0119+11 & FSRQ & I/II & 1.8 & 28.9 & 20.4 & 10.7 & 0.34\\
0133+476 & DA 55 & FSRQ & II & 2.2 & 24.1 & 17.0 & 7.4 & 0.28\\
0202+149 & 4C +15.05 & FSRQ & II & 2.0 & 25.8 & 17.8 & 5.3 & 0.18\\
0202+319 & B2 0202+31 & FSRQ & II & 3.1 & 16.6 & 11.4 & 14.3 & 0.77\\
0212+735 & S5 0212+73 & FSRQ & II & 4.7 & 10.9 & 7.5 & 6.6 & 0.54\\
0215+015 & OD 026 & FSRQ & II & 1.3 & 41.1 & 28.8 & 20.1 & 0.44\\
0219+428 & 3C 66A & ISP BL Lac & I/II & 3.4 & 16.2 & 21.5 & 20.8 & 1.2\\
0224+671 & 4C +67.05 & FSRQ & I/II & 2.3 & 22.0 & 15.3 & 9.8 & 0.40\\
0234+285 & 4C +28.07 & FSRQ & II & 1.6 & 32.7 & 23.8 & 11.3 & 0.32\\
0241+622 & 7C 0241+6215 & FSRQ & I & - & - & - & 8.0 & -  \\
0250-225 & OD -283 & FSRQ & II & 4.3 & 11.7 & 8.0 & 14.8 & 1.12\\
0300+470 & 4C +47.08 & LSP BL Lac & I/II & 5.1 & 11.0 & 14.6 & 25.5 & 2.2\\
0333+321 & NRAO 140 & FSRQ & I/II & 2.5 & 20.6 & 14.4 & 8.3 & 0.37\\
0336-019 & CTA 26 & FSRQ & I/II & 1.4 & 36.6 & 26.5 & 5.9 & 0.15\\
0355+508 & NRAO 150 & FSRQ & I & - & - & - & 2.8 & -  \\
0403-132 & PKS 0403-13 & FSRQ & I/II & 1.6 & 32.2 & 22.8 & 5.5 & 0.15\\
0420-014 & PKS 0420-01 & FSRQ & II & 5.2 & 9.7 & 6.6 & 8.9 & 0.81\\
0422+004 & PKS 0422+00 & LSP BL Lac & I & - & - & - & * & -  \\
0430+052 & 3C 120 & FSRQ & II & 4.7 & 10.8 & 7.4 & 2.6 & 0.21\\
0440-003 & NRAO 190 & FSRQ & I & - & - & - & 2.6 & - \\
0446+112 & PKS 0446+11 & FSRQ & II & 4.5 & 11.2 & 7.7 & 22.2 & 1.75\\
0454-234 & PKS 0454-234 & FSRQ & II & 5.1 & 10.1 & 6.9 & 42.6 & 3.75\\
0458-020 & S3 0458-02 & FSRQ & II & 2.3 & 21.9 & 15.2 & 5.0 & 0.21\\
0528+134 & PKS 0528+134 & FSRQ & II & 1.8 & 28.3 & 19.5 & 10.3 & 0.32\\
0529+075 & OG 050 & FSRQ & II & 1.8 & 28.6 & 20.0 & 28.5 & 0.90\\
0529+483 & TXS 0529+483 & FSRQ & II & 1.6 & 31.7 & 22.1 & 10.9 & 0.31\\
0539-057 & PKS 0539-057 & FSRQ & II & 4.0 & 12.8 & 8.8 & 10.9 & 0.75\\
0552+398 & DA 193 & FSRQ & I & - & - & - & 24.0 & -  \\
0605-085 & OC -010 & FSRQ & II & 1.8 & 29.2 & 20.9 & 11.0 & 0.35\\
0607-157 & PKS 0607-15 & FSRQ & I & - & - & - & 4.3 & - \\
0642+449 & OH 471 & FSRQ & II & 3.6 & 14.1 & 9.7 & 7.8 & 0.49\\
0716+714 & S5 0716+714 & LSP BL Lac & II & 1.8 & 28.9 & 20.5 & 10.2 & 0.32\\
0723-008 & PKS 0723-008 & LSP BL Lac & I & - & - & - & 12.1 & -  \\
0730+504 & TXS 0730+504 & FSRQ  & I/II & 2.3 & 22.1 & 15.6 & 7.2 & 0.30\\
0735+178 & OI 158 & FSRQ & I/II & 6.0 & 8.5 & 5.8 & 8.8 & 0.92\\
0736+017 & OI 061 & FSRQ & II & 2.3 & 22.0 & 15.3 & 9.3 & 0.38\\
0738+313 & OI 363 & FSRQ & II & 2.9 & 17.6 & 12.1 & 4.2 & 0.21\\
0742+103 & PKS B0742+103 & GPS1 Quasar  & II & 10.8 & 4.7 & 3.3 & 9.5 & 1.79\\
0745+241 & S3 0745+24 & FSRQ & I/II & 4.7 & 10.9 & 7.5 & 8.3 & 0.68\\
0748+126 & OI 280 & FSRQ & II & 2.2 & 23.7 & 16.3 & 9.3 & 0.35\\
0754+100 & PKS 0754+100 & ISP BL Lac & I/II & 2.2 & 23.6 & 16.2 & 11.8 & 0.45\\
0804+499 & OJ 508 & FSRQ & I & - & - & - & 11.7 & -  \\
0805-077 & PKS 0805-07 & FSRQ & II & 0.9 & 59.7 & 44.5 & 6.3 & 0.10\\
0808+019 & OJ 014 & LSP BL Lac & II & 2.3 & 22.0 & 15.3 & 12.3 & 0.50\\
0823+033 & PKS 0823+033 & LSP BL Lac & I/II & 2.5 & 20.5 & 14.3 & 10.6 & 0.47\\
0827+243 & OJ 248 & FSRQ & II & 1.6 & 31.7 & 22.1 & 13.5 & 0.38\\
0829+046 & OJ 049 & LSP BL Lac & I/II & 3.1 & 16.6 & 11.4 & 10.7 & 0.57  \\
0834-201 & PKS 0834-20 & FSRQ & II & 4.3 & 11.7 & 7.9 & 23.2 & 1.75\\
0836+710 & 4C +71.07 & FSRQ & II & 1.6 & 32.3 & 23.1 & 4.9 & 0.14\\
0838+133 & 3C 207 & FSRQ & II & 2.9 & 17.7 & 12.1 & 2.2 & 0.11\\
0851+202 & OJ 287 & ISP BL Lac & I/II & 2.2 & 23.9 & 16.8 & 13.4 & 0.51\\
0859-140 & PKS B0859-140 & CSS Quasar & II & 4.9 & 10.4 & 7.0 & 4.4 & 0.37\\
0906+015 & 4C +01.24 & FSRQ & II & 1.4 & 35.6 & 24.7 & 3.0 & 0.08\\
0917+449 & S4 0917+44 & FSRQ & II & 17.7 & 3.0 & 2.1 & 9.1 & 2.77\\
0917+624 & OK 630 & FSRQ & II & 2.7 & 19.1 & 13.4 & 9.2 & 0.44\\
0923+392 & 4C +39.25 & FSRQ & II & 10.8 & 4.7 & 3.3 & * & *\\
0945+408 & 4C +40.24 & FSRQ & II & 1.6 & 31.9 & 22.4 & 6.3 & 0.18\\
0954+658 & S4 0954+65 & FSRQ & II & 2.5 & 20.4 & 14.2 & 16.9 & 0.74\\
0955+476 & OK 492 & FSRQ & II & 5.1 & 10.1 & 6.9 & 16.0 & 1.41 \\
1011+496 & 1ES 1011+496 & HSP BL Lac  & I & - & - & - & 5.7 & -  \\
1036+054 & PKS 1036+054 & FSRQ & II & 5.2 & 9.7 & 6.6 & 9.1 & 0.83\\
1038+064 & 4C +06.41 & FSRQ & II & 2.9 & 17.6 & 12.1 & 2.1 & 0.10\\
1045-188 & PKS 1045-18 & FSRQ & II & 2.9 & 17.5 & 12.0 & 2.8 & 0.14\\
1055+018 & 4C +01.28 & FSRQ & I/II & 3.8 & 13.4 & 9.2 & 10.1 & 0.67\\
1101+384 & Mrk 421 & HSP BL Lac  & I & - & - & - & 10.4 & -\\
1118-056 & PKS 1118-05 & FSRQ & I & - & - & - & 9.7 & -\\
1127-145 & PKS 1127-14 & FSRQ & I/II & 2.2 & 23.8 & 16.6 & 2.7 & 0.10\\
1150+497 & 4C +49.22 & FSRQ & II & 1.8 & 28.4 & 19.6 & 8.2 & 0.26\\
1150+812 & S5 1150+81 & FSRQ & II & 3.1 & 16.6 & 11.4 & 7.4 & 0.40\\
1156+295 & 4C +29.45 & FSRQ & II & 1.4 & 36.7 & 26.6 & 13.8 & 0.35\\
1215+303 & ON 325 & HSP BL Lac  & I & - & - & - & 7.0 & - \\
1219+044 & 4C +04.42 & FSRQ & I & - & - & - & 5.2 & - \\
1219+285 & W Comae & ISP BL Lac & I/II & 3.4 & 14.9 & 10.3 & 8.1 & 0.49\\
1222+216 & 4C +21.35 & FSRQ & II & 1.3 & 41.3 & 29.2 & 3.4 & 0.07\\
1226+023 & 3C 273 & FSRQ & II & 2.2 & 23.8 & 16.6 & 3.7 & 0.14\\
1236+049 & BZQ J1239+0443 & FSRQ & II & 8.5 & 6.0 & 4.1 & 15.3 & 2.25\\
1244-255 & PKS 1244-255 & FSRQ & II & 9.4 & 5.4 & 3.8 & 21.5 & 3.50\\
1253-055 & 3C 279 & FSRQ & I/II & 1.6 & 32.1 & 22.7 & 3.2 & 0.09\\
1302-102 & PG 1302-102 & FSRQ & II & 3.2 & 15.7 & 10.7 & 6.7 & 0.38\\
1308+326 & OP 313 & FSRQ & II & 1.3 & 41.7 & 29.9 & 8.8 & 0.19\\
1329-049 & OP -050 & FSRQ & II & 4.0 & 12.8 & 8.8 & 7.4 & 0.51\\
1334-127 & PKS 1335-127 & FSRQ & II & 2.0 & 26.0 & 18.1 & 6.1 & 0.21\\
1406-076 & PKS B1406-076 & FSRQ & II & 1.3 & 42.1 & 30.7 & 7.3 & 0.16\\
1413+135 & PKS B1413+135 & LSP BL Lac & I & - & - & - & 3.9 & -  \\
1417+385 & B3 1417+385 & FSRQ & II & 2.2 & 24.1 & 17.0 & * & *\\
1418+546 & OQ 530 & LSP BL Lac & I/II & 6.7 & 7.6 & 5.2 & 7.1 & 0.83\\
1458+718 & 3C 309.1 & CSS Quasar & II & 4.7 & 10.9 & 7.5 & 2.3 & 0.19\\
1502+106 & OR 103 & FSRQ & II & 1.8 & 28.4 & 19.6 & 22.3 & 0.70\\
1504-166 & PKS 1504-167 & FSRQ & II & 7.6 & 6.7 & 4.6 & 7.2 & 0.95\\
1510-089 & PKS 1510-08 & FSRQ & II & 1.3 & 41.9 & 30.3 & 9.5 & 0.21\\
1514-241 & AP Librae & LSP BL Lac & I/II & 4.7 & 10.8 & 7.3 & 8.5 & 0.70\\
1520+319 & B2 1520+31 & FSRQ & I & - & - & - & 28.2 & -  \\
1538+149 & 4C +14.60 & LSP BL Lac & I/II & 3.6 & 14.2 & 9.8 & 10.9 & 0.69\\
1546+027 & PKS 1546+027 & FSRQ & II & 2.7 & 19.1 & 13.4 & 8.2 & 0.39\\
1548+056 & 4C +05.64 & FSRQ & II & 2.7 & 18.9 & 13.0 & 6.5 & 0.31\\
1551+130 & OR 186 & FSRQ & II & 3.6 & 14.2 & 9.8 & 5.8 & 0.36\\
1606+106 & 4C +10.45 & FSRQ & II & 1.8 & 29.1 & 20.8 & 8.8 & 0.28\\
1611+343 & DA 406 & FSRQ & II & 1.3 & 42.4 & 31.3 & 8.4 & 0.18\\
1622-297 & PKS 1622-29 & FSRQ & II & 1.8 & 28.9 & 20.5 & 5.2 & 0.17\\
1633+382 & 4C +38.41 & FSRQ & II & 1.3 & 42.4 & 31.3 & 6.3 & 0.14\\
1637+574 & OS 562 & FSRQ & II & 2.3 & 21.9 & 15.2 & 6.1 & 0.25\\
1641+399 & 3C 345 & FSRQ & II & 1.8 & 29.2 & 21.0 & 10.1 & 0.32\\
1642+690 & 4C +69.21 & FSRQ & I/II & 2.2 & 23.6 & 16.3 & 7.0 & 0.26\\
1652+398 & Mrk 501 & HSP BL Lac  & I & - & - & - & 8.2 & - \\
1655+077 & PKS 1655+077 & FSRQ & II & 2.2 & 23.8 & 16.5 & 2.3 & 0.09\\
1700+685 & TXS 1700+685 & FSRQ & II & 4.0 & 12.9 & 8.9 & 7.5 & 0.52\\
1726+455 & S4 1726+45 & FSRQ & II & 13.0 & 4.0 & 2.8 & 8.9 & 1.99\\
1730-130 & NRAO 530 & FSRQ & II & 1.3 & 41.7 & 29.8 & 3.7 & 0.08\\
1749+096 & 4C +09.57 & LSP BL Lac & II & 4.0 & 12.9 & 8.9 & 10.1 & 0.70\\
1751+288 & B2 1751+28 & FSRQ & II & 7.8 & 6.5 & 4.5 & 7.8 & 1.05\\
1758+388 & B3 1758+388B & FSRQ & I/II & 13.5 & 3.8 & 2.7 & 8.8 & 2.06\\
1800+440 & S4 1800+44 & FSRQ & II & 2.2 & 24.1 & 17.0 & 6.2 & 0.23\\
1803+784 & S5 1803+784 & LSP BL Lac & I/II & 2.9 & 17.7 & 12.2 & 13.4 & 0.68\\
1807+698 & 3C 371 & ISP BL Lac & I & - & - & - & 4.3 & -  \\
1823+568 & 4C +56.27 & LSP BL Lac & I/II & 1.3 & 41.1 & 28.9 & 3.8 & 0.08\\
1828+487 & 3C 380 & CSS Quasar & I/II & 2.5 & 20.6 & 14.5 & 5.4 & 0.24\\
1846+322 & B2 1846+32A & FSRQ & II & 4.5 & 11.3 & 7.7 & 17.6 & 1.38\\
1849+670 & S4 1849+67 & FSRQ & I/II & 1.4 & 36.1 & 25.4 & 11.4 & 0.29\\
1908-201 & PKS B1908-201 & FSRQ & II & 6.9 & 7.4 & 5.1 & 24.5 & 2.93\\
1928+738 & 4C +73.18 & FSRQ & II & 3.8 & 13.5 & 9.2 & 5.2 & 0.34\\
1936-155 & PKS 1936-15 & FSRQ & I/II & 5.6 & 9.0 & 6.2 & 8.2 & 0.80\\
2005+403 & TXS 2005+403 & FSRQ & I/II & 3.1 & 16.5 & 11.3 & 7.5 & 0.40\\
2007+777 & S5 2007+77 & ISP BL Lac & I & - & - & - & 8.2 & -  \\
2008-159 & PKS 2008-159 & FSRQ & II & 6.1 & 8.2 & 5.6 & 4.6 & 0.49\\
2013+370 & MG2 J201534+3710 & FSRQ & I/II & 2.5 & 20.3 & 14.0 & 9.4 & 0.41\\
2021+317 & 4C +31.56 & Unknown & II & 8.5 & 6.0 & 4.1 & 17.2 & 2.53\\
2022-077 & PKS 2023-07 & FSRQ & II & 1.4 & 36.1 & 25.5 & 15.9 & 0.40\\
2037+511 & 3C 418 & FSRQ & II & 7.9 & 6.4 & 4.4 & 9.2 & 1.27\\
2121+053 & OX 036 & FSRQ & II & 2.7 & 18.9 & 13.1 & 12.5 & 0.59\\
2128-123 & PKS 2128-12 & FSRQ & I/II & 5.1 & 10.0 & 6.9 & 2.6 & 0.23\\
2131-021 & 4C -02.81 & FSRQ & I/II & 1.6 & 31.8 & 22.1 & 8.1 & 0.23\\
2134+004 & PKS 2134+004 & FSRQ & II & 6.0 & 8.5 & 5.8 & 4.0 & 0.41\\
2136+141 & OX 161 & FSRQ & II & 7.2 & 7.0 & 4.8 & 20.5 & 2.57\\
2141+175 & OX 169 & FSRQ & II & 12.1 & 4.2 & 3.0 & 11.8 & 2.47\\
2145+067 & 4C +06.69 & FSRQ & II & 10.6 & 4.8 & 3.3 & 10.2 & 1.89\\
2155-152 & PKS 2155-152 & FSRQ & II & 1.8 & 28.7 & 20.1 & 8.5 & 0.27\\
2200+420 & BL Lacertae & ISP BL Lac & I/II & 3.1 & 16.5 & 11.3 & 17.1 & 0.92\\
2201+171 & PKS 2201+171 & FSRQ & I/II & 1.8 & 28.5 & 19.8 & 10.0 & 0.31\\
2201+315 & 4C +31.63 & FSRQ & II & 3.8 & 13.5 & 9.3 & 6.6 & 0.43\\
2209+236 & PKS 2209+236 & FSRQ & I & - & - & - & 10.0 & -\\
2216-038 & PKS 2216-03 & FSRQ & II & 4.5 & 11.3 & 7.7 & 3.7 & 0.29\\
2223-052 & 3C 446 & FSRQ & II & 1.6 & 32.0 & 22.5 & 6.8 & 0.19 \\
2227-088 & PHL 5225 & FSRQ & II & 15.3 & 3.4 & 2.4 & 4.4 & 1.16\\
2230+114 & CTA 102 & FSRQ & II & 3.6 & 14.2 & 9.8 & 6.3 & 0.4\\
2243-123 & PKS 2243-123 & FSRQ & II & 5.8 & 8.8 & 6.0 & 6.7 & 0.68\\
2251+158 & 3C 454.3 & FSRQ & II & 2.3 & 22.0 & 15.3 & 8.9 & 0.36\\
2254+074 & PKS 2254+074 & LSP BL Lac & I & - & - & - & 10.5 & -  \\
2331+073 & TXS 2331+073 & FSRQ & II & 5.6 & 9.0 & 6.2 & 12.8 & 1.25\\
2344+514 & 1ES 2344+514 & HSP BL Lac & I & - & - & - & 10.0 & - \\
2345-167 & PKS 2345-16 & FSRQ & II & 2.7 & 18.8 & 12.9 & 14.4 & 0.68\\
2351+456 & 4C +45.51 & FSRQ & II & 1.6 & 32.5 & 23.4 & 14.9 & 0.42\\
\end{longtable}
{* : Jets with undefined aperture angle ($R^2 <0.1$).}

\end{document}